\documentclass[showpacs,preprintnumbers,amsmath,amssymb,appendix]{revtex4}
\newcommand{\be}{\begin{equation}}
\newcommand{\ee}{\end{equation}}
\newcommand{\bea}{\begin{eqnarray}}
\newcommand{\eea}{\end{eqnarray}}
\newcommand{\ba}{\begin{array}}
\newcommand{\ea}{\end{array}}

\usepackage{hyperref}
\usepackage{rotating}

\begin{document}

\title{Spin $\frac{1}{2}{^+}$, spin $\frac{3}{2}{^+}$ and
transition magnetic moments of low lying and charmed baryons}
\author{Neetika Sharma$^a$, Harleen Dahiya$^a$, P.K. Chatley$^a$ and Manmohan Gupta$^b$}
\affiliation{$^a$Department of Physics, Dr. B.R. Ambedkar National
Institute of Technology, Jalandhar, 144011, India
\\ $^b$Department of Physics, Centre of Advanced Study in Physics, Panjab
University, Chandigarh 160014, India}
\date{\today}

\begin{abstract}
Magnetic moments of the low lying and charmed spin $\frac{1}{2}{^+}$
and spin $\frac{3}{2}{^+}$ baryons have been calculated in the SU(4)
chiral constituent quark model ($\chi$CQM) by including the
contribution from $c \bar c$ fluctuations. Explicit calculations
have been carried out for the contribution coming from the valence
quarks, ``quark sea'' polarizations and their orbital angular
momentum. The implications of such a model have also been studied
for magnetic moments of the low lying spin $\frac{3}{2}{^+} \to
\frac{1}{2}{^+}$ and $\frac{1}{2}{^+} \to \frac{1}{2}{^+}$
transitions as well as the transitions involving charmed baryons.
The predictions of $\chi$CQM not only give a satisfactory fit for
the baryons where experimental data is available but also show
improvement over the other models. In particular, for the case of
$\mu(p)$, $\mu(\Sigma^{+})$, $\mu(\Xi^{0})$, $\mu(\Lambda)$,
Coleman-Glashow sum rule for the low lying spin $\frac{1}{2}{^+}$
baryons and $\mu(\Delta{^+})$, $\mu(\Omega^{-})$ for the low lying
spin $\frac{3}{2}{^+}$ baryons, we are able to achieve an excellent
agreement with data. For the spin $\frac{1}{2}{^+}$ and spin
$\frac{3}{2}{^+}$ charmed baryon magnetic moments, our results are
consistent with the predictions of the QCD sum rules, Light Cone sum
rules and Spectral sum rules. For the cases where ``light'' quarks
dominate in the valence structure, the sea and orbital contributions
are found to be fairly significant however, they cancel in the right
direction to give the correct magnitude of the total magnetic
moment. On the other hand, when there is an excess of ``heavy''
quarks, the contribution of the ``quark sea'' is almost negligible,
for example, $\mu(\Omega_{c}^{0})$, $\mu(\Lambda_{c}^{+})$,
$\mu(\Xi_{c}^{+})$, $\mu(\Xi_{c}^{0})$, $\mu(\Omega_{cc}^{+})$,
$\mu(\Omega^{-})$, $\mu(\Omega_{c}^{*0})$, $\mu(\Omega_{cc}^{*+})$
and $\mu(\Omega_{ccc}^{*++})$. The effects of configuration mixing
and quark masses have also been investigated.

\end{abstract}
\maketitle

\section{Introduction}\label{intro}

The possible size of {\it intrinsic} charm (IC) content of the
nucleon \cite{ic} has been estimated to understand the
phenomenological implications of the presence of heavy quarks in the
nucleon. The heavy flavor charmed baryons play an important role to
understand the dynamics of light quarks in the bound state as well
as to understand QCD at the hadronic scale \cite{garcilazo}. On the
other hand, the static and electromagnetic properties like masses,
magnetic moments etc. give valuable information regarding the
internal structure of baryons \cite{pdg} in the nonperturbative
regime. Since there is no direct experimental data on the IC
content, one has to resort to the nucleon models to obtain
information on its contribution.

The magnetic moments of spin $\frac{1}{2}{^+}$, spin
$\frac{3}{2}{^+}$ charmed baryons and their transition magnetic
moments have been considered in different approaches in literature,
however, none of the phenomenological models is able to give a
complete description. Calculations based on different realizations
of spin-flavor symmetries have been done in the non-relativistic
quark model (NRQM) \cite{choudhury} which have been further extended
to incorporate the confinement \cite{rlpm}, chiral symmetry with
exchange currents \cite{ccm} and Poincare covariance \cite{rqm}. The
charmed baryons magnetic moments have been investigated in the
Skyrme model \cite{skyrme} and the bound state approach \cite{bsa}
considering the heavy baryons as heavy mesons bound in the field of
light baryons. Recently, the charmed baryons magnetic moments have
been calculated in the relativistic three-quark model where the
internal quark structure of baryons is modeled by three-quark
currents \cite{rtqm}. More recently, magnetic moments have been
studied by considering the effective mass of quark bound inside the
baryon \cite{patel}. The magnetic moments of spin $\frac{1}{2}{^+}$
and spin $\frac{3}{2}{^+}$ including transition magnetic moments of
charmed baryons have been also investigated in QCD sum rules (QCDSR)
\cite{wanglee}, QCD Spectral sum rules (QSSR) \cite{qssr} and light
cone QCD sum rules (LCQSR) \cite{lcqsr,tam3/2,tam}.

It would be important to mention here that the {\it intrinsic} heavy
quarks are created from the quantum fluctuations associated with the
bound state hadron dynamics and the process is completely determined
by nonperturbative mechanisms \cite{charm}. Recently, it has been
shown that one of the important model which finds application in the
nonperturbative regime of QCD is the chiral constituent quark model
($\chi$CQM) \cite{wein,eichten,cheng}. The $\chi$CQM with spin-spin
generated configuration mixing \cite{dgg,isgur} is able to give the
satisfactory explanation for the spin and flavor distribution
functions \cite{hd,song}, strangeness content of the nucleon
\cite{hds}, weak vector and axial-vector form factors \cite{ns}
etc.. When coupled with the ``quark sea'' polarization, orbit
angular momentum of the ``quark sea'' (referred as Cheng and Li
Mechanism) and confinement effects it is able to give a excellent
fit to the octet and decuplet baryon magnetic moments and a perfect
fit to the violation of Coleman-Glashow sum rule
\cite{cgsr,cheng1,johan,hdorbit}. The quantum fluctuations generated
by broken chiral symmetry in $\chi$CQM should be able to provide a
viable estimate of the heavier quark flavor, for example, $c \bar
c$, $b \bar b$ and $t \bar t$. However, it is known that these
flavor fluctuations are much suppressed in the case of $b \bar b$
and $t \bar t$ as compared to the $c \bar c$ because the intrinsic
heavy quark contributions scale as $1/M^2_q$, where $M_q$ is the
mass of the heavy quark \cite{charm,chengspin}. In this context, the
scope of model was extended to the broken SU(4) symmetry which
successfully predicted the important role played by IC content in
determining the spin and flavor structure of the nucleon
\cite{songcharm,hdcharm}. In the light of above investigations, it
becomes desirable to broaden the scope of Ref. \cite{hdcharm} by
extending the calculations to magnetic moments and transition
magnetic moments of the charmed baryons.

The purpose of the present paper is to formulate in detail the
magnetic moments of spin $\frac{1}{2}{^+}$ and spin
$\frac{3}{2}{^+}$ charmed baryons in the SU(4) framework of
$\chi$CQM. The magnetic moments of the low lying spin
$\frac{3}{2}{^+} \to \frac{1}{2}{^+}$ and $\frac{1}{2}{^+} \to
\frac{1}{2}{^+}$ transitions as well as the transitions involving
charmed baryons would also be calculated. The generalized Cheng-Li
mechanism has been incorporated to calculate explicitly the
contribution coming from the valence spin polarization, ``quark
sea'' polarization and its orbital angular momentum. In order to
understand the implications of charm quarks in the baryons without
any valence charm quarks and to make our analysis more responsive,
we would also like to calculate the low lying octet and decuplet
baryons magnetic moments in the SU(4) framework of $\chi$CQM.
Further, it would also be interesting to examine the effects of the
configuration mixing, symmetry breaking parameters, confinement
effects, quark masses etc. on the magnetic moments.

The plan of work is as follows. To facilitate discussion, in Sec.
\ref{cccm}, SU(4) $\chi$CQM is revisited with an emphasis on the
details of spin dynamics. In Sec. \ref{magmom}, we first present the
essential details of Cheng-Li mechanism to obtain the magnetic
moments of baryons and in the subsequent subsections, we discuss the
baryon magnetic moments with spin $\frac{1}{2}{^+}$, spin
$\frac{3}{2}{^+}$ and their transition magnetic moments,
respectively. Few typical cases pertaining to charmed baryons have
been worked out in detail in each case. Discussion on the various
inputs used in the analysis is presented in Sec. \ref{inputs}. In
Sec. \ref{results}, we present our numerical results and their
comparison with the other model predictions. Finally, we summarize
our results in Sec. \ref{summary}. The details pertaining to the
wave functions for charmed baryons have been presented in the
Appendix A.

\section{Spin structure in chiral constituent quark model}
\label{cccm} The basic process in the $\chi$CQM \cite{wein} is the
internal emission of a Goldstone Boson (GB) by a constituent quark
which further splits into a $q \bar q$ pair as $q_{\pm} \rightarrow
{\rm GB}^{0}+q'_{\mp} \rightarrow (q \bar q^{'})+ q'_{\mp}\,,$ where
$q \bar q^{'} +q^{'}$ constitutes the ``quark sea''
\cite{cheng,hd,song,johan}. The details of $\chi$CQM in the SU(4)
framework have already been discussed in literature
\cite{songcharm,hdcharm}, however, for the sake of readability of
manuscript, we discuss here the essentials of the spin structure of
the baryons used in the calculations of magnetic moments.

The effective Lagrangian describing interaction between quarks and
sixteen GBs, consisting of 15-plet and a singlet, can be expressed
as ${{\cal L}}= g_{15}{\bf \bar q}\left(\Phi\right) {\bf q}$, where
$g_{15}$ is the coupling constant and GBs field $\Phi$ is { \bea
\Phi = \left( \ba{ccccc} \frac{\pi^0}{\sqrt 2} +
\beta\frac{\eta}{\sqrt 6}+ \zeta\frac{\eta^{'}}{4\sqrt 3}-
\gamma\frac{\eta_c}{4} & \pi{^+} & \alpha K{^+} & \gamma \bar{D}^0\\
\pi^- & - \frac{\pi^0}{\sqrt 2} + \beta \frac{\eta}{\sqrt 6}
+\zeta\frac{\eta^{'}}{4\sqrt 3} - \gamma\frac{\eta_c}{4}& \alpha K^0
& \gamma D^-\\ \alpha K^- & \alpha \bar{K}^0 &- \beta
\frac{2\eta}{\sqrt 6} + \zeta\frac{\eta^{'}}{4\sqrt 3}-
\gamma\frac{\eta_c}{4} & \gamma D^-_s\\ \gamma D^0  &\gamma D{^+}&
\gamma D^+_s& - \zeta\frac{3\eta^{'}}{4\sqrt 3} + \gamma
\frac{3\eta_c}{4} \ea \right)\,. \eea} SU(4) symmetry breaking is
introduced by considering $M_c>M_s>M_{u,d}$ as well as by
considering the masses of GBs to be nondegenerate $(M_{\eta_{c}}>
M_{\eta^{'}}> M_{K,\eta}>M_{\pi})$ \cite{cheng,songcharm,hdcharm}.
The parameter $a(=|g_{15}|^2)$ denotes the transition probability of
chiral fluctuation of the splitting $u(d) \rightarrow d(u)+
\pi^{+(-)}$, whereas $a \alpha^2$, $a \beta^2$, $a \zeta^2$ and $a
\gamma^2$ denote the probabilities of transitions of $u(d)
\rightarrow s + K^{-(o)}$, $u(d,s) \rightarrow u(d,s) + \eta$,
$u(d,s) \rightarrow u(d,s) + \eta^{'}$ and  $u(d) \rightarrow c
+\bar{D}^0(D^-)$, respectively.

The spin structure of the baryon is defined as \cite{cheng,hd,johan}
\be \widehat B \equiv \langle B|{\cal N}|B \rangle\,, \label{bnb}
\ee where $|B\rangle$ is the baryon wave function and ${\cal N}$ is
the number operator defined as \be {\cal N}=n_{u_{+}}u_{+}+
n_{u_{-}}u_{-} + n_{d_{+}}d_{+} + n_{d_{-}}d_{-} + n_{s_{+}}s_{+} +
n_{s_{-}}s_{-} + n_{c_{+}}c_{+} + n_{c_{-}}c_{-}\,, \label{number}
\ee $n_{q_{\pm}}$ being the number of $q_{\pm}$ quarks. The valence
spin polarizations ($\Delta q_{{\rm val}}= q_{+}- q_{-}$) for a
given baryon can be calculated using the spin and flavor wave
functions detailed in Appendix A. The ``quark sea'' spin
polarizations ($\Delta q_{{\rm sea}}$) can be calculated by
substituting for each valence quark \be q_{\pm}\rightarrow \sum P_q
q_{\pm}+ |\psi(q_{\pm})|^2\,, \label{qpq} \ee where $\sum P_q$ is
the probability of emission of GBs from a $q$ quark and
$|\psi(q_{\pm})|^2$ is the probability of transforming a $q_{\pm}$
quark \cite{hdcharm}.

\section{Magnetic moments in $\chi$CQM}
\label{magmom} The magnetic moment of a given baryon receives
contributions from the valence quarks, ``quark sea'' and orbital
angular momentum of the ``quark sea'' following Cheng and Li
\cite{cheng,hds,cheng1,hdorbit} and is expressed as \be \mu(B)_{{\rm
total}} = \mu(B)_{{\rm val}} + \mu(B)_{{\rm sea}} + \mu (B)_{{\rm
orbit}} \,, \label{totalmag} \ee where $\mu(B)_{{\rm val}}$ and
$\mu(B)_{{\rm sea}}$ represent the contributions of the valence
quarks and the ``quark sea'' to the magnetic moments due to spin
polarizations. The term $\mu(B)_{{\rm orbit}}$ corresponds to the
orbital angular momentum contribution of the ``quark sea''.

In terms of quark magnetic moments and spin polarizations, the
valence, sea and orbital contributions can be defined as \be
\mu(B)_{{\rm val}} = \sum_{q=u,d,s,c}{\Delta q_{{\rm
val}}\mu_q}\,,~~ \mu(B)_{{\rm sea}} = \sum_{q=u,d,s,c}{\Delta
q_{{\rm sea}}\mu_q}\,,~~ \mu(B)_{{\rm orbit}} = \sum_{q=u,d,s,c}
{\Delta q_{{\rm val}}~\mu(q_{+} \rightarrow {q}_{-}^{'})}
\,,\label{mag} \ee where $\mu_q = \frac{e_q}{2 M_q}$ ($q=u,d,s,c$)
is the quark magnetic moment, $\mu(q_{+} \rightarrow {q}_{-}^{'})$
is the orbital moment for any chiral fluctuation, $e_q$ and $M_q$
are the electric charge and the mass respectively for the quark $q$.

The valence and sea quark spin polarizations ($\Delta q_{{\rm val}}$
and $\Delta q_{{\rm sea}}$) for a given baryon can be calculated as
discussed in the previous section and Ref. \cite{hdorbit}. The
orbital angular momentum contribution of each chiral fluctuation is
given as \cite{cheng1,hdorbit} \be \mu(q_{+} \rightarrow
{q}_{-}^{'}) = \frac{e_{q^{'}}}{2M_q} \langle l_q \rangle +
\frac{{e}_{q} - {e}_{q^{'}}}{2 {M}_{{\rm GB}}}\langle {l}_{{\rm GB}}
\rangle \label{qq} \,, \ee where $\langle l_q \rangle =
\frac{{M}_{{\rm GB}}}{M_q + {M}_{{\rm GB}}}$ and $\langle l_{{\rm
GB}} \rangle = \frac{M_q}{M_q + {M}_{{\rm GB}}}\label{lq}$. The
quantities ($l_q$, $l_{{\rm GB}}$) and ($M_q$, ${M}_{{\rm GB}}$) are
the orbital angular momenta and masses of quark and GBs,
respectively. The orbital moment of each process in Eq.(\ref{qq}) is
then multiplied by the probability for such a process to take place
to yield the magnetic moment due to all the transitions starting
with a given valence quark. The details of the orbital moment in the
SU(3) framework has already been worked out in Ref. \cite{hdorbit}.
In this work, we extend our calculations to include the contribution
from $c \bar c$ fluctuations. For example, \be [\mu(u_{\pm}
\rightarrow )]=\pm a \left [ \left(\frac{1}{2} +\frac{\beta^2}{6}+
\frac{\zeta^2}{48}+ \frac{\gamma^2}{16} \right) \mu (u_{+}
\rightarrow u_-)+ \mu (u_{+}\rightarrow d_-)+ \alpha^2 \mu (u_+
\rightarrow s_-)+ \gamma^2 \mu (u_+ \rightarrow c_-)\right ] \,,
\label{muu} \ee \be [\mu (d_{\pm} \rightarrow )] = \pm a \left [ \mu
(d_{+} \rightarrow u_-)+ \left (\frac{1}{2}+ \frac{\beta^2}{6} +
\frac{\zeta^2}{48}+ \frac{\gamma^2}{16} \right) \mu (d_{+}
\rightarrow  d_-)+ \alpha^2 \mu (d_+ \rightarrow s_-)+ \gamma^2 \mu
(d_+ \rightarrow c_-) \right ] \,, \label{mud} \ee \be [\mu (s_{\pm}
\rightarrow )]= \pm a \left [\alpha^2 \mu (s_{+} \rightarrow u_-) +
\alpha^2 \mu (s_+ \rightarrow d_-) + \left (\frac{2}{3} \beta^2+
\frac{\zeta^2}{48} + \frac{\gamma^2}{16} \right) \mu (s_{+}
\rightarrow s_- ) +\gamma^2 \mu (s_+ \rightarrow c_-)\right ] \,,
\label{mus} \ee and \be [\mu (c_{\pm} \rightarrow )]= \pm a \left
[\gamma^2 \mu (c_{+} \rightarrow u_-) + \gamma^2 \mu (c_{+}
\rightarrow d_-)+ \gamma^2 \mu (c_{+} \rightarrow s_-) +\left(
\frac{3}{16}\zeta^2+ \frac{9}{16}\gamma^2 \right) \mu (c_{+}
\rightarrow c_- ) \right ] \,. \label{muc} \ee The above equations
can easily be generalized by including the coupling breaking and
mass breaking terms. The orbital moments of $u$, $d$, $s$ and $c$
quarks in terms of the $\chi$CQM parameters ($a, \alpha, \beta,
\zeta, \gamma$), quark masses ($M_u,M_d,M_s,M_c$) and GB masses
($M_{\pi},M_{k},M_{\eta},M_{\eta'},M_{D},M_{D_s},M_{\eta_c}$), are
respectively given as \bea [\mu(u_+ \rightarrow)] &=& a \left
[\frac{3 M^2_{u}}{2 {M}_{\pi}(M_u+ {M}_{\pi})}-
\frac{\alpha^2(M^2_{K}- 3 M^2_{u})}{2 {M}_{K}(M_u+ {M}_{K})} +
\frac{ \gamma^2 M_D}{(M_u+ M_D)} \right. \nonumber \\ && \left .+
\frac{\beta^2 M_{\eta}}{6(M_u+ {M}_{\eta})}+ \frac{\zeta^2
M_{\eta'}}{48(M_u+ {M}_{\eta'})} + \frac{\gamma^2
M_{\eta_c}}{{16}(M_u+ {M}_{\eta_c})} \right] {\mu}_N \,, \\
\label{orbitu} [\mu(d_+ \rightarrow)] &=& a \frac{M_u}{M_d}\left
[\frac{3( M^2_{\pi}-2 M^2_{d})}{4 {M}_{\pi}(M_d+ {M}_{\pi})}-
\frac{\alpha^2 M_{K}}{2(M_d+ {M}_{K})}+ \frac{\gamma^2(2 M^2_{D}- 3
M^2_{d})} {2 {M}_{D}(M_d+ {M}_{D})} \right. \nonumber \\ && \left. -
\frac{\beta^2 M_{\eta}}{12(M_d+ {M}_{\eta})}- \frac{\zeta^2
M_{\eta'}}{96(M_d+ {M}_{\eta'})}- \frac{\gamma^2 M_{\eta_c}}{32(M_d+
{M}_{\eta_c})} \right ] {\mu}_N \,, \label{orbitd} \eea \be[\mu(s_+
\rightarrow)] = a \frac{M_u}{M_s}\left[ \frac{\alpha^2 (M^2_{K}-3
M^2_s)}{2{M}_{K}(M_s+ {M}_{K})} + \frac{\gamma^2( 2 M^2_{D_s}-3
M^2_s)}{2M_{D_s}(M_s+ M^2_{D_s})} - \frac{\beta^2 M_{\eta}}{3(M_s+
{M}_{\eta})}- \frac{\zeta^2 M_{\eta'}}{96(M_s+ {M}_{\eta'})} -
\frac{\gamma^2 M_{\eta_c}}{32(M_s+{M}_{\eta_c})} \right ]{\mu}_N\,,
\label{orbits} \ee  \be[\mu(c_+ \rightarrow)] = a
\frac{M_u}{M_c}\left[\frac{\gamma^2( M^2_{D}+ 3 M^2_c)}{2M_{D}(M_c+
M^2_{D})}- \frac{\gamma^2(M^2_{D_s}-3 M^2_c)}{2 M_{D_s}(M_c+
M^2_{D_s})}+ \frac{3 \zeta^2 M_{\eta'}}{16 (M_c+ {M}_{\eta'})} +
\frac{9 \gamma^2 M_{\eta_c}}{16(M_c+ {M}_{\eta_c})} \right
]{\mu}_N\,, \label{orbitc} \ee where $\mu_N$ is the nuclear
magneton.

After discussing the general formalism to calculate the valence,
sea and orbital contributions to the magnetic moments, we now
discuss the explicit calculations for the low lying and charmed
spin $\frac{1}{2}{^+}$ and spin $\frac{3}{2}{^+}$ baryons as well
as their transition magnetic moments.

\renewcommand{\thesubsection}{3.\arabic{subsection}}
\subsection{Magnetic moments of spin $\frac{1}{2}{^+}$ baryons}
\label{secspin1/2} The magnetic moments of all the spin
$\frac{1}{2}{^+}$ baryons can be calculated using
Eq.(\ref{totalmag}). The spin structure of a spin $\frac{1}{2}{^+}$
baryon (from Appendix A) is given as \be \hat B \equiv \langle B
|{\cal N}|B\rangle= {\cos}^2 \phi {\langle 120,{^2}20_M|{\cal
N}|120,{^2}20_M\rangle}_B+ {\sin}^2 \phi {\langle 168,{^2}20_M|{\cal
N}|168,{^2}20_M\rangle}_B\,. \label{spinst} \ee In this section, as
an example, we detail the calculations of the one single and one
double charmed baryon $\Xi_{c}^{'+}$ and $\Xi_{cc}^{++}$.

The valence spin structure for the single charmed baryon
$\Xi_{c}^{'+}$ can be expressed as \bea \widehat{\Xi}_{c}^{'+} =
{\cos}^2 \phi \left (\frac{5}{6} u_{+}+ \frac{1}{6} u_{-}+
\frac{5}{6} s_{+} +\frac{1}{6} s_{-}+ \frac{1}{3} c_{+}+ \frac{2}{3}
c_{-}\right)+ {\sin}^2 \phi \left( \frac{2}{3} u_{+}+ \frac{1}{3}
u_{-}+ \frac{2}{3} s_{+}+ \frac{1}{3} s_{-}+ \frac{2}{3} c_{+} +
\frac{1}{3}c_{-} \right)\,, \label{xic} \eea leading to the valence
contribution to the magnetic moment of $\Xi_{c}^{'+}$ \be
\mu(\Xi_{c}^{'+})_{{\rm val}}= {\cos}^2 \phi \left(\frac{2}{3}
\mu_u+ \frac{2}{3}\mu_s- \frac{1}{3}\mu_c\right)+ {\sin}^2 \phi
\left(\frac{1}{3}\mu_u+ \frac{1}{3}\mu_s+ \frac{1}{3}\mu_c
\right)\,. \label{valxic} \ee The spin structure of the ``quark
sea'' ($\Delta q_{{\rm sea}}$) can be calculated by substituting
Eq.(\ref{qpq}) for every valence quark in Eq.(\ref{xic}), leading to
the ``quark sea'' contribution to the magnetic moment of
$\Xi_{c}^{'+}$ expressed as \bea \mu(\Xi_{c}^{'+})_{{\rm sea}} &=&
-\frac{a}{3}{\cos}^2 \phi \left[ \left(4+ 4 \alpha^2+
\frac{2}{3}\beta^2+ \frac{\zeta^2}{12}+
\frac{9}{8}\gamma^2\right)\mu_{u}+ \left(2+ 2\alpha^2-
\gamma^2\right)\mu_{d}+ \left(6\alpha^2+ \frac{8}{3}\beta^2+
\frac{\zeta^2}{12}+ \frac{9}{8}\gamma^2 \right)\mu_{s} \right.
\nonumber \\ && \left.  - \frac{1}{8}\left(3\zeta^2+
\gamma^2\right)\mu_{c} \right] - \frac{a}{3}{\sin}^2 \phi \left[
\left(2+ 2\alpha^2+ \frac{\beta^2}{3}+ \frac{\zeta^2}{24}+
\frac{33}{16} \gamma^2 \right)\mu_{u}+ \left(1+ \alpha^2+ \gamma^2
\right)\mu_{d} \right. \nonumber \\ && \left.  + \left(3 \alpha^2+
\frac{4}{3}\beta^2+ \frac{\zeta^2}{24}+
\frac{33}{16}\zeta^2\right)\mu_{s}+ \frac{1}{8}\left(3\zeta^2+
49\gamma^2 \right)\mu_{c} \right]\,. \label{seaxic} \eea

The orbital contribution of the ``quark sea'' to the total
magnetic moment of $\Xi_{c}^{'+}$, obtained using Eqs.(\ref{mag})
and (\ref{valxic}), can be expressed as \bea
\mu(\Xi_{c}^{'+})_{{\rm orbit}} = {\cos}^2 \phi \left(
\frac{2}{3}\mu (u_+ \rightarrow) + \frac{2}{3}\mu (s_+
\rightarrow)- \frac{1}{3}\mu (c_+ \rightarrow)\right)+ {\sin}^2
\phi \left( \frac{1}{3}\mu (u_+ \rightarrow)+ \frac{1}{3}\mu
(s_+\rightarrow)+\frac{1}{3}\mu (c_+ \rightarrow)\right)\,.
\label{orbitxic} \eea Substituting the valence, sea and orbital
contribution from Eqs.(\ref{valxic}), (\ref{seaxic}) and
(\ref{orbitxic}) in Eq.(\ref{totalmag}), we can calculate the
total magnetic moment of $\Xi_{c}^{'+}$.

For the double charmed baryon $\Xi_{cc}^{++}$, the valence spin
structure can be expressed as \be \widehat{\Xi}_{cc}^{++}= {\cos}^2
\phi \left( \frac{1}{3}u_{+}+ \frac{2}{3}u_{-}+ \frac{5}{3}c_{+}+
\frac{1}{3} c_{-}\right)+ {\sin}^2 \phi\left( \frac{2}{3}u_{+}+
\frac{1}{3}u_{-}+ \frac{4}{3}c_{+}+ \frac{2}{3}c_{-}
\right)\,,\label{xicc} \ee giving the valence, sea and orbital
contribution to the magnetic moment of $\Xi_{cc}^{++}$ as \bea
\mu(\Xi_{cc}^{++})_{{\rm val}} &=& {\cos}^2 \phi \left(-
\frac{1}{3}\mu_u+ \frac{4}{3}\mu_c \right)+ {\sin}^2 \phi
\left(\frac{1}{3}\mu_u+ \frac{2}{3}\mu_{c} \right)\,,\\
\label{valxicc}   \mu(\Xi^{++}_{cc})_{{\rm sea}} &=&
\frac{a}{3}{\cos}^2 \phi \left[\left(2+ \alpha^2+ \frac{\beta^2}{3}+
\frac{\zeta^2}{24}- \frac{47}{16} \gamma^2\right)\mu_{u}+ \left(1- 4
\gamma^2 \right)\mu_{d}+ \left (\alpha^2- 4\gamma^2 \right) \mu_{s}-
\frac{1}{2}\left(3\zeta^2+ 31\gamma^2 \right)\mu_{c} \right]
\nonumber \\ &-& \frac{a}{3}{\sin}^2 \phi \left[ \left(2+ \alpha^2+
\frac{\beta^2}{3}+ \frac{\zeta^2}{24}+ \frac{49}{16} \gamma^2
\right)\mu_{u}+ \left(1+ 2\gamma^2\right)\mu_{d}+ \left( \alpha^2+
2\gamma^2\right)\mu_{s}+ \frac{1}{4}\left(3\zeta^2+
37\gamma^2\right)\mu_{c} \right]\,, \\ \label{seaxicc}
\mu(\Xi_{cc}^{++})_{{\rm orbit}} &=& {\cos}^2 \phi
\left(-\frac{1}{3}\mu (u_+ \rightarrow)+ \frac{4}{3}\mu (c_+
\rightarrow) \right)+ {\sin}^2 \phi \left( \frac{1}{3}\mu (u_+
\rightarrow)+ \frac{2}{3}\mu (c_+ \rightarrow) \right)\,.
\label{orbitxicc} \eea The valence, sea and orbital contribution
from Eqs.(\ref{valxicc}), (\ref{seaxicc}) and (\ref{orbitxicc}) give
the total magnetic moment of $\Xi_{cc}^{++}$. Similarly, one can
calculate the valence, sea and orbital contributions to the magnetic
moments of all the spin $\frac{1}{2}{^+}$ baryons. The expressions
for the valence and sea contributions to the magnetic moments of the
low lying and charmed spin $\frac{1}{2}^+$ baryons have been
presented in the Table \ref{spin1/2charm}.

\subsection{Magnetic moments of spin $\frac{3}{2}{^+}$ baryons}
\label{secspin3/2} In this section, we detail the calculations of
magnetic moments of spin $\frac{3}{2}{^+}$ baryons by taking the
example of a charmed baryon $\Xi_c^{*+}$. From Appendix A, the spin
structure of a spin $\frac{3}{2}{^+}$ baryon is given as \be
\widehat B^{*} \equiv \langle B^{*}|{\cal N}|B^{*}\rangle ={\langle
120, {^4}20_S|{\cal N}|120, {^4}20_S\rangle}_{B^{*}} \,.
\label{spinst3/2} \ee The valence spin structure of $\Xi_c^{*+}$ can
be expressed as \be \widehat{\Xi}_{c}^{*+} = u_{+}+ s_{+}+ c_{+} \,,
\label{decval} \ee giving the valence contribution to the magnetic
moment as \be \mu(\Xi_{c}^{*+})_{{\rm val}}= \mu_{u}+ \mu_{s}+
\mu_{c}\,. \label{decvalxic} \ee The ``quark sea'' contribution to
the magnetic moment of $\Xi_{c}^{*0}$ can be calculated by
substituting Eq.(\ref{qpq}) for every valence quark in
Eq.(\ref{decval}), giving the sea contribution as \bea
\mu(\Xi_{c}^{*+})_{{\rm sea}}&=&-a\left[\left(2+ 2 \alpha^2+
\frac{\beta^2}{3}+ \frac{\zeta^2}{24}+ \frac{33}{16}\gamma^2
\right)\mu_{u}+ \left(1+ \alpha^2+ \gamma^2 \right)\mu_{d}\right.
\nonumber \\&+& \left. \left(3\alpha^2+ \frac{4}{3}\beta^2+
\frac{\zeta^2}{24}+ \frac{33}{16}\gamma^2 \right)\mu_{s}+
\frac{1}{8}\left(3\zeta^2+ 49\gamma^2\right) \mu_{c}\right]\,.
\label{decseaxic} \eea The orbital angular momentum contribution to
the magnetic moment of $\Xi_{c}^{*+}$ is given as \be
\mu(\Xi_{c}^{*+})_{{\rm orbit}}=\mu (u_+ \rightarrow)+ \mu (s_+
\rightarrow)+ \mu (c_+ \rightarrow) \,. \label{decorbitxic} \ee
Substituting the valence, sea and orbital contribution from
Eqs.(\ref{decvalxic}), (\ref{decseaxic}) and (\ref{decorbitxic}) in
Eq.(\ref{totalmag}), we can calculate the total magnetic moment of
$\Xi_{c}^{*+}$. The valence, sea and orbital contribution to the
magnetic moments of other spin $\frac{3}{2}{^+}$ charmed baryons can
similarly be calculated and the expressions for the valence and sea
contribution to the total magnetic moment of the spin
$\frac{3}{2}^+$ charmed baryons have been presented in Table
\ref{spin3/2}.

\subsection{Transition magnetic moments}
\label{spin3/2tospin1/2} In this section, we calculate the
transition magnetic moments for the radiative decays $B_i
\rightarrow B_f + \gamma,$ where $B_i$ and $B_f$ are the initial and
final baryons, for the spin $\frac{3}{2}{^+} \to \frac{1}{2}{^+}$
and $\frac{1}{2}{^+} \to \frac{1}{2}{^+}$ transitions of the
baryons. In particular, the transition magnetic moments considered
in this work are for spin $\frac{3}{2}{^+} \to \frac{1}{2}{^+}$
transitions corresponding to the charmless decuplet to octet
transitions (10 $\to$ 8), single charmed sextet to anti-triplet
transitions ($6 \to \bar 3$), single charmed sextet to sextet
transitions ($6 \to 6$) and double charmed triplet to triplet
transitions ($3 \to 3$) transitions. On the other hand, the spin
$\frac{1}{2}{^+} \to \frac{1}{2}{^+}$ transitions considered are for
the charmless octet to octet transitions ($8 \to 8$) and single
charmed anti-triplet to sextet transitions ($\bar 3 \to 6$). The
details of the structure have been presented in Appendix A.

The transition magnetic moment can be calculated from the matrix
element \be \widehat{B_{i}B_f}(\textbf{k})=\langle B_f,
J_z=\frac{1}{2}|{\cal N} e^{-\iota k.z}| B_{i}, J_z=\frac{1}{2}
\rangle \,,\ee where $\textbf{k}$ is the momentum of the photon. As
an example, we discuss here the case of transition magnetic moment
of the $6 \to 6$ transition ($\Xi^{*+}_{c}\Xi^{'+}_{c}$). The spin
structure for the ($\Xi^{*+}_{c}\Xi^{'+}_{c}$) transition is given
as \be \widehat{\Xi^{*+}_{c}\Xi^{'+}_{c}}(\textbf{k})=
\frac{\sqrt{2}}{3}(-u_+- s_+ + 2c_+)\cdot e^{-\frac{1}{6}k^2R^2} \,,
\ee giving the valence contribution to the magnetic moment of
($\Xi^{*+}_{c}\Xi^{'+}_{c}$) transition as \be
\mu(\Xi^{*+}_{c}\Xi^{'+}_{c})_{\rm val}= \frac{\sqrt{2}}{3}(-
\mu_{u}- \mu_{s}+ 2\mu_{c})\cdot e^{-\frac{1}{6}k^2R^2}
\,.\label{deltaval} \ee The ``quark sea'' contribution can be
calculated by making substitution Eq.({\ref{qpq}) for every valence
quark. The quark sea contribution for the magnetic moment of
($\Xi^{*+}_{c}\Xi^{'+}_{c}$) transition is then expressed as {\bea
\mu(\Xi^{*+}_{c}\Xi^{'+}_{c})_{{\rm sea}}& =& \frac{\sqrt{2}}{3}a
\left[ \left(2+ 2\alpha^2+ \frac{\beta^2}{3}+ \frac{\zeta^2}{24}-
\frac{15}{16} \gamma^2\right)\mu_{u}+ \left(1+ \alpha^2-
2\gamma^2\right)\mu_{d} \right. \nonumber \\&+& \left.
\left(3\alpha^2+ \frac{4}{3}\beta^2+ \frac{\zeta^2}{24}-
\frac{15}{16}\gamma^2\right)\mu_{s}- \frac{1}{4}\left(3\zeta^2+
25\gamma^2\right)\mu_{c} \right] \cdot e^{-\frac{1}{6}k^2 R^2} \,.
\label{deltasea} \eea} The orbital angular momentum contribution in
this case is \be \mu (\Xi^{*+}_{c}\Xi^{'+}_{c})_{{\rm orbit}}=
\frac{\sqrt{2}}{3}\bigg(- \mu(u_+\rightarrow)- \mu(s_+\rightarrow+
2\mu(c_+\rightarrow ) \bigg)\cdot e^{-\frac{1}{6}k^2R^2}
\,.\label{deltaorbit} \ee The total magnetic moment for the
transition ($\Xi^{*+}_{c}\Xi^{'+}_{c}$) can be calculated by adding
Eqs.(\ref{deltaval}), (\ref{deltasea}) and (\ref{deltaorbit}). The
detailed expressions for the valence, sea and orbital contribution
to the magnetic moments for all other transitions can be calculated
similarly and the expressions  are presented in the Table
\ref{trans}.

\section{Input parameters} \label{inputs}
In this section, we discuss the various input parameters needed for
the numeric calculation of the magnetic moments of spin
$\frac{1}{2}^+$ and spin $\frac{3}{2}^+$ baryons. The valence, sea
and orbital contributions to the magnetic moment in $\chi$CQM with
SU(4) broken symmetry involve the symmetry breaking parameters and
mixing angle $\phi$. The symmetry breaking parameters $a$, $a
\alpha^2$, $a \beta^2$, $a \zeta^2$, $a \gamma^2$ representing
respectively, the probabilities of fluctuations of a constituent
quark into pions, $K$, $\eta$, $\eta^{'}$, $\eta_c$, are expected to
follow the hierarchy $a> a \alpha^2 > a \beta^2> a \zeta^2 >a
\gamma^2$ as they are dominated by the mass differences. As a
consequence, the probability of emitting a heavier meson such as $D$
from a lighter quark is much smaller than that of emitting the light
meson such as $K$, $\eta$ and $\eta{'}$ etc.. The symmetry breaking
parameters are usually fixed by the spin polarization functions
$\Delta u$, $\Delta d$ and $Q^2$ independent parameter
$\Delta_3(=\Delta u-\Delta d)$ \cite{emc,adams,ellis} as well as the
flavor distribution functions $\bar u-\bar d$ and $\bar u/\bar d$
\cite{nmc,e866}, measured from the deep inelastic scattering
experiments. The mixing angle $\phi$ is fixed by fitting neutron
charge radius \cite{yaouanc}. A fine grained analysis with the
symmetry breaking lead to the following set of symmetry breaking
parameters as the best fit \be a=0.12\,,~~ \alpha \simeq \beta=
0.45\,,~~ \zeta = -0.21 \,, ~~ {\rm and} ~~ \gamma=0.11\,. \nonumber
\ee

In addition to the parameters of $\chi$CQM and mixing angle $\phi$
as discussed above, the orbital angular momentum contributions are
characterized by the quark, GB masses and the harmonic-oscillator
radius parameter $R$. For evaluating their contribution, we have
used their on shell mass values in accordance with several other
similar calculations \cite{mpi1,isgur}. For the constituent quark
masses $u$, $d$, $s$, $c$, we have used their widely accepted values
in hadron spectroscopy $M_u = M_d = 0.33$ GeV, $M_s = 0.51$ GeV,
$M_c = 1.70$ GeV. The quark masses and corresponding magnetic
moments have to be further adjusted by the quark confinement effects
\cite{hdorbit,effm}. For the low lying baryons, Kerbikov {\it et
al.} \cite{conf} have given a successful description of the magnetic
moment with confinement effects playing a leading role. However, in
the present case the simplest way to incorporate this adjustment
\cite{effm} is to first express $M_q$ in the magnetic moment
operator in terms of $M_B$, the mass of the baryon obtained
additively from the quark masses, which then is replaced by $M_B +
\Delta M$, $\Delta M$ being the mass difference between the
experimental value and $M_B$. This leads to the following
adjustments in the quark magnetic moments: $\mu_{u}=2[1-(\Delta
M/M_B)] {\mu}_N$, $\mu_{d} = -[1-(\Delta M/M_B)] {\mu}_N$, $\mu_{s}
= -M_u/M_s [1-(\Delta M/M_B)]{\mu}_N$ and $\mu_c =
2M_u/M_c[1-(\Delta M/M_B)]{\mu}_N$.

\section{Results and Discussion}
\label{results} The parameters discussed above have been used to
calculate the various spin polarization functions, non-singlet
components $\Delta_3$ and $\Delta_8$ and flavor distribution
functions in SU(4) $\chi$CQM. The values obtained for the case of
proton are as follows \be \Delta u=0.93 \,, \Delta d=-0.34 \,,
\Delta s=-0.03 \,, \Delta c=-0.002 \,, \Delta_3=1.2696 \,, \Delta_8=
0.64 \,, \ee \be \bar u=0.23\,, \bar d=0.34\,, \bar s=0.086\,, \bar
c=0.005\,, \bar u-\bar d= -0.11\,, \frac{\bar d}{\bar u}=1.49\,. \ee
We find that a fairly good fit is achieved in the parameters listed
above when compared with the latest data \cite{pdg,charm,adams}. In
particular, the agreement corresponding to the strangeness and
intrinsic charm contribution to the nucleon in terms of the
magnitude as well as the sign is quite satisfactory when compared
with the latest data \cite{pdg,charm,adams}. A detailed
implications of these parameters have already been discussed in Ref.
\cite{hdcharm}. It is interesting to mention here that these
strangeness and charm related parameters have not been taken as
inputs in our calculations and still a satisfactory agreement is
obtained. In addition, SU(4) $\chi$CQM leads to many new predictions
on observables which are directly related to the charm content of
nucleon and are found to be almost an order of magnitude smaller
than the strange quark contributions but not entirely insignificant.
Consistency of these charm related parameters can be checked by
future experiments.

The spin polarization functions discussed above have been used to
calculate the baryon magnetic moments. In Tables \ref{spin1/2num}
and \ref{spin3/2num}, we have presented the results for the magnetic
moments of low lying and charmed spin $\frac{1}{2}^+$, spin
$\frac{3}{2}^+$ baryons. In Table \ref{transnum}, we have presented
the magnetic moments for the low lying spin $\frac{3}{2}{^+} \to
\frac{1}{2}{^+}$ and $\frac{1}{2}{^+} \to \frac{1}{2}{^+}$
transitions as well as the transitions involving charmed baryons. In
the tables, we have presented the explicit results for the valence,
sea and orbital contributions to the magnetic moments. We have also
compared our results with the predictions of NRQM \cite{choudhury},
Lattice QCD \cite{lattice} and recent experimental data available
\cite{pdg}. Since there is no experimental information available for
charmed baryon magnetic moments, we have presented the predictions
of QCD sum rules (QCDSR) \cite{wanglee}, QCD Spectral sum rules
(QSSR) \cite{qssr}, Light Cone QCD sum rules (LCQSR)
\cite{lcqsr,tam3/2,tam}.

A cursory look at the tables reveal that the our results are smaller
than the NRQM predictions in most of the cases and our results are
not only in agreement with available data but also show improvement
over other models in most of the cases where the experimental data
is available. On the other hand, for the case of the magnetic
moments where experimental data is not available, our results are
consistent with the results of QCDSR, QSSR, LCQSR as well as with
the other models existing in literature. One can also observe that
the orbital part contributes with the same sign as valence quark
distribution, while the sea part contribute with the opposite sign
making the sea and orbital contributions significant. The sum of
residual ``quark sea'' and valence quark contribution give the
magnetic moment of baryons.

From Table \ref{spin1/2num}, when we compare our results for the
spin $\frac{1}{2}{^+}$ baryons with the available experimental data
as well as the other model calculations, we find that our model is
able to get a fairly good account of the most of magnetic moments,
wherever the experimental data is available. Presently, experimental
information is available for the low lying octet baryons and
violation of Coleman-Glashow sum rule ($\Delta$CG) \cite{cgsr}. It
is interesting to observe that our results for the magnetic moments
of ${p}$, ${\Sigma^{+}}$, ${\Xi^{0}}$ and ${\Lambda}$ give a perfect
fit to the experimental values \cite{pdg} whereas for all other
octet baryons our predictions are within 10\% of the observed
values. Besides this, we have also been able to get an excellent fit
to $\Delta$CG. The fit becomes all the more impressive when it is
realized that none of the magnetic moments are used as inputs and
$\Delta$CG can be described without resorting to additional
parameters.

A closer look at the table reveals that if an attempt is made to
explain the contribution of the orbital angular momentum of the
``quark sea'', we find the contribution of orbital angular momentum
to be as important as that of the ``quark sea'' contribution through
the spin polarization of the $q \bar q$ pairs. In fact, the sea and
orbital contributions are fairly significant as compared to the
valence contributions and they cancel in the right direction giving
the right magnitude of the total magnetic moment. For example, the
valence contributions of $p$, $\Sigma^{+}$ and $\Xi^{0}$ are higher
in magnitude than the experimental value but the sea contribution
being higher in magnitude than the orbital contribution reduces the
valence contribution leading to a better agreement with data.
Similarly, in the case of $n$, $\Sigma^{-}$ and $\Xi^{-}$ the
valence contribution in magnitude is lower than the experimental
value but in these cases the sea contribution is lower than the
orbital part so it adds on to the valence contribution again
improving agreement with data. It is important to mention here that
the IC contribution to the proton spin polarizations and hence
magnetic moments is quite small so the predictions of the SU(4)
$\chi$CQM do not differ significantly from our earlier results in
the SU(3) $\chi$CQM for the octet baryons \cite{hdorbit}.

In the case of charmed baryons also, there is a significant
contribution from the ``quark sea'' spin and orbital angular
momentum. Only in the case of $\Omega_{c}^{0}$, $\Lambda_{c}^{+}$,
$\Xi_{c}^{+}$, $\Xi_{c}^{0}$ and $\Omega_{cc}^{+}$, the magnetic
moment if dominated by the valence contribution as the sea and
orbital contributions are quite small in magnitude. This is because
of the fact that the above mentioned baryons are dominated by the
``heavy'' quarks in the valence structure. Thus, in a very
interesting manner, the orbital and sea contributions together add
on to the valence contributions leading to better agreement with
data. This not only endorses the earlier conclusion of Cheng and Li
\cite{cheng1} but also suggests that the Cheng-Li mechanism could
perhaps provide the dominant dynamics of the constituents in the
nonperturbative regime of QCD on which further corrections could be
evaluated.

From Table \ref{spin3/2num}, we can compare our results for the low
lying as well as charmed spin $\frac{3}{2}{^+}$ baryons with other
model calculations as well as with the available experimental data.
In this case also, we have presented the explicit results for the
valence, sea and the orbital contributions. For the magnetic moments
of the low lying decuplet baryons, only three experimental results
are presently available. Our predicted value for $\Delta^{++}=4.51$,
is well within the experimental range $3.7 \sim 7.5$ \cite{pdg}.
Similarly, for the case of $\Delta{^+}$ and $\Omega^{-}$ our
predicted values 2.0 and $-$1.71, agree with the experimentally
observed values ($2.7^{+1.0}_{-1.3} \pm 1.5 \pm 3$ \cite{kotulla}
and $-1.94 \pm 0.31$ \cite{diehl}, respectively). For all other
baryons our predictions are consistent with the predictions of the
QCDSR \cite{wanglee}, LCQSR \cite{tam3/2}, Lattice QCD
\cite{lattice} and other models existing in literature. However,
there is a small discrepancy in the case of $\Sigma^{*0}$ magnetic
moment while comparing our results with other model calculations. In
this case, the contribution of the orbital part is negligible and
the valence and sea contributions are of the same order. The valence
and sea contribution being of opposite signs cancel each other
completely leading to a very small $\Sigma^{*0}$. Any experimental
data on $\Sigma^{*0}$ would have important implications for the
Cheng-Li mechanism. For the charmed spin $\frac{3}{2}{^+}$ baryons,
since there is no experimental information available, we have
compare our results with the predictions of the LCQSR \cite{tam3/2}.
Our results are consistent with their predictions and also with the
other models existing in the literature \cite{patel,wanglee}.

On the closer scrutiny of the results we find that in the cases
where there is an excess of up and down quarks in the valence
structure, the contribution of the ``quark sea'' and its orbital
angular momentum is quite significant when compared with the valence
contribution. On the other hand, when there is an excess of strange
and charm quarks in the valence structure, the contribution of the
``quark sea'' and its orbital angular momentum is almost negligible
as compared to the valence contribution. This can be easily
understood when we compare the sea and orbital contributions of
$\Omega^{-}$, $\Omega_{c}^{*0}$, $\Omega_{cc}^{*+}$ and
$\Omega_{ccc}^{*++}$ with the sea and orbital contributions of the
other baryons. In these cases, the total magnetic moment is more or
less the same as the valence contribution whereas in all other cases
there is a significant contribution from the resultant sea and
orbital contributions. It would be interesting to mention here that
this is due to the fact that the strange and charm contribution to
the magnetic moment is almost an order of magnitude smaller than the
up and down quarks thus leading to a very small contribution from
the ``heavy'' quarks when compared with the contribution coming from
the ``light'' quarks.

In Table \ref{transnum}, we have presented results for the magnetic
moments of the spin $\frac{3}{2}{^+} \to \frac{1}{2}{^+}$
transitions corresponding to the charmless decuplet to octet
transitions (10 $\to$ 8), single charmed sextet to anti-triplet
transitions ($6 \to \bar 3$), single charmed sextet to sextet
transitions ($6 \to 6$) and double charmed triplet to triplet
transitions ($3 \to 3$) transitions. We have also presented the
results for the spin $\frac{1}{2}{^+} \to \frac{1}{2}{^+}$
transitions corresponding to the charmless octet to octet
transitions (8 $\to$ 8) and single charmed anti-triplet to sextet
transitions ($\bar 3 \to 6$). Experimental data is available for
only the low lying 8 $\to$ 8 transition ($\Sigma^0 \to \Lambda +
\gamma$). Our prediction for this decay is 1.60 ($1.61 \pm 0.08$
\cite{pdg}). There is no experimental data available for any other
charmed baryons transition magnetic moments as well as for the other
low lying spin $\frac{3}{2}{^+} \to \frac{1}{2}{^+}$ transitions so
we have presented the predictions of LCQSR \cite{tam} and Lattice
QCD \cite{lattice}, wherever the results are available. For the
magnetic moment of the $({\Delta \to p + \gamma})$ transition, an
empirical estimate can be made from the helicity amplitudes
$A_{\frac{1}{2}}$ = $-$ 0.135 $\pm$ 0.005 GeV$^{-\frac{1}{2}}$, and
$A_{\frac{3}{2}}$ = $-$ 0.250 $\pm$0.008 GeV$^{-\frac{1}{2}}$
\cite{pdg} as inputs in the decay rate and the magnetic moment
extracted is $\mu({\Delta p})$ = 3.46 $\pm$ 0.03 $\mu_N$
\cite{tiator}. The magnetic moment of $\mu(\Delta p)$ transition is
a long standing problem and most of the approaches in literature
underestimate it. Our predicted value 2.87 $\mu_N$ is below the
experimental results. The implications of $\chi$CQM and Cheng-Li
mechanism perhaps can be substantiated by future measurements of
$\mu({\Delta p})$.

Implications of configuration mixing, quark masses and confinement
effects have also been investigated. In the spin $\frac{1}{2}{^+}$
baryon magnetic moments, it is found that the inclusion of Cheng-Li
mechanism predicts the results in the right direction even when
configuration mixing is not included, however, the inclusion of
confinement effects alongwith configuration mixing plays a crucial
role in fitting the individual magnetic moments. Interestingly, we
find that the masses $M_u = M_d = 330$ MeV, after corrections due to
configuration mixing and confinement effects, provide the best fit
for the magnetic moments. This implies a deeper significance for the
$\chi$CQM coupling breaking and the quark masses parameters
employed.

\section{Summary and conclusion} \label{summary}
To summarize, in order to enlarge the scope of SU(4) chiral
constituent quark model ($\chi$CQM) and to estimate the
phenomenological contribution of $c \bar c$ fluctuations, we have
carried out a detailed analysis  of the magnetic moments of the low
lying and charmed spin $\frac{1}{2}{^+}$ and spin $\frac{3}{2}{^+}$
baryons as well as of their transitions. Using the generally
accepted values of the quark masses, the parameters of $\chi$CQM
have been fixed from the latest data pertaining to $\bar u-\bar d$
asymmetry and spin polarization functions, the explicit
contributions coming from the valence quarks, the ``quark sea''
contribution as well as its orbital angular momentum through the
generalized Cheng-Li mechanism have been calculated.

For the low lying $\frac{1}{2}{^+}$ and spin $\frac{3}{2}{^+}$
baryons where experimental data is available, the $\chi$CQM
predictions not only give a satisfactory fit but also show
improvement over the other models. In particular, for the case of
$\mu(p)$, $\mu(\Sigma^{+})$, $\mu(\Xi^{0})$, $\mu(\Lambda)$,
violation of Coleman-Glashow sum rule for the spin $\frac{1}{2}{^+}$
baryons and $\mu(\Delta{^+})$, $\mu(\Omega^{-})$ for the spin
$\frac{3}{2}{^+}$ baryons, we are able to achieve an excellent
agreement with data. For all the other low lying octet and decuplet
baryons our predictions are within 10\% of the observed values. For
the spin $\frac{1}{2}{^+}$ and spin $\frac{3}{2}{^+}$ charmed baryon
magnetic moments, our results are very much in agreement with recent
theoretical estimates. It is observed that the orbital part
contributes with the same sign as valence quark distribution, while
the sea part contribute with the opposite sign. Further, for the
cases where ``light'' quarks dominate in the valence structure, the
resultant sea and orbital contributions are found to be fairly
significant as compared to the valence contributions.  On the other
hand, when there is an excess of ``heavy'' quarks, the contribution
of the ``quark sea'' is almost negligible, for example,
$\mu(\Omega_{c}^{0})$, $\mu(\Lambda_{c}^{+})$, $\mu(\Xi_{c}^{+})$,
$\mu(\Xi_{c}^{0})$, $\mu(\Omega_{cc}^{+})$, $\mu(\Omega^{-})$,
$\mu(\Omega_{c}^{*0})$, $\mu(\Omega_{cc}^{*+})$ and
$\mu(\Omega_{ccc}^{*++})$. However, it is interesting that the sea
and orbital parts cancel in the right direction to give the correct
magnitude of the total magnetic moment.

The implications of such a model have also been been studied for the
case of low lying spin $\frac{3}{2}{^+} \to \frac{1}{2}{^+}$
transition magnetic moments as well as for the $\frac{1}{2}{^+} \to
\frac{1}{2}{^+}$ transitions involving charmed baryons. In this case
also, the contribution of orbital angular momentum is found to be as
important as that of the spin polarization of the $q \bar q$ pairs.
Implications of configuration mixing and quark masses have also been
investigated. Interestingly, we find that generalized Cheng-Li
mechanism coupled with corrections due to configuration mixing and
confinement effects, provide the best fit for the magnetic moments.
This suggests that constituent quarks and weakly interacting
Goldstone bosons provide the appropriate degree of freedom in the
nonperturbative regime of QCD. This fact can perhaps can be
substantiated by a measurement of the magnetic moments of charmed
baryons. Several groups BTeV, SELEX Collaboration are contemplating
the possibility of performing it in the near future.

\vskip .2cm
{\bf ACKNOWLEDGMENTS}\\
H.D. would like to thank Department of Science and Technology,
Government of India, for financial support.

\appendix
\section{The wave function convention for the baryon}
\renewcommand{\theequation}{A-\arabic{equation}}
\setcounter{equation}{0} The total wave function for the three quark
system made from any of the $u$, $d$, $s$ or $c$ quarks is given as
$|{SU(8)}\otimes{O(3)} \rangle = \varphi \chi \psi$, where $\varphi$
is a flavor wave function, $\chi$ is a spin wave function and $\psi$
is a spatial wave function. The $SU(8)$ multiplet is decomposed into
$SU(4) \otimes SU(2)$ flavor and spin multiplets, respectively. The
multiplet numerology for the subset of baryons belonging to SU(4)
flavor multiplets, is $4\times 4 \times 4$ =
20${_S}$+20${_{M}}$+20${_{M}}$+$\bar 4$, where the symmetry 20-plet
consists of 10+6+3+1 and the mixed symmetry 20-plet consists of
8+6+$\bar 3$+3 baryons flavor states. For the details of the
definition of spatial part of the wave function ($\psi^{s},
\psi^{'}, \psi^{''})$ represented by the ${O(3)}$, we refer the
reader to reference \cite{yaoubook}.

In order to understand the structure of charmed baryon wave
functions and sign conventions used in this work, we present here
the $SU(4) \otimes SU(2)$ content of the $SU(8)$ multiplet which
is given as \bea 120 &\supset&{^4}20_S+{^2}20_M \,,\nonumber\\ 168
&\supset&{^2}20_S+{^4}20_M+{^2}20_M+{^2}\bar 4 \,,\nonumber\\ 56
&\supset& {^2}\bar 4+{^2}20_M \,. \eea The $SU(8)\otimes O(3)$
wave functions for the spin $\frac{1}{2}{^+}$ and
$\frac{3}{2}{^+}$ baryons are respectively, \bea |B \rangle \simeq
|120, {^2}20_M\rangle_{N=0}&=& \frac{1}{\sqrt 2}(\chi^{'}
\varphi^{'} + \chi^{''} \varphi^{''}) \psi^{s}(0^+) \,,
\nonumber\\ |B^{*}\rangle \simeq |120, {^4}20_S{\rangle_{N=0}}&=&
\chi^{s} \varphi^s \psi^s(0^+) \,.\label{3/2}\eea

To incorporate the effect of configuration mixing generated by the
spin-spin interactions \cite{hd,hdorbit,hdcharm} which has been
shown to improve the prediction of the $\chi$CQM, the complete wave
function for the spin $\frac{1}{2}{^+}$ baryons can be expressed as
\bea |B \rangle &=& {\cos} \phi |120, {^2}20_M\rangle_{N=0} + {\sin}
\phi|168,{^2}20_M\rangle_{N=2}\,, \label{config} \eea where $|168,
{^2}20_M\rangle_{N=2} = \frac{1}{2}(( \varphi^{'}\chi^{''}+
\varphi^{''} \chi^{'})\psi^{'}(0^+) +(\varphi^{'} \chi^{'} -
\varphi^{''} \psi^{''}) \psi^{''}(0^+)) \,.$ The explicit flavor
wave functions for the spin $\frac{1}{2}{^+}$ baryons are as follows
{\bea \ba{cccc} \hline &{\rm Baryon} & \varphi{'} & \varphi{''}\\
\hline (8,0)& p & \frac{1}{\sqrt 2}(udu-duu)&\frac{1}{\sqrt
6}(2uud-udu-duu)\nonumber\\&n &\frac{1}{\sqrt
2}(udd-dud)&\frac{1}{\sqrt 6}(dud+udd-2ddu) \nonumber\\
&\Sigma^{+} &\frac{1}{\sqrt 2}(usu-suu) & \frac{1}{\sqrt
6}(2uus-suu-usu) \nonumber\\&\Sigma^{0} &
\frac{1}{2}(sud+sdu-usd-dsu)& \frac{1}{2\sqrt
3}(sdu+dsu+sud+usd-2uds-2dus)\nonumber\\&\Sigma^-&\frac{1}{\sqrt
2}(sdd-dsd)& \frac{1}{\sqrt 6}(2dds-sdd-dsd) \nonumber\\&\Lambda
&\frac{1}{2\sqrt 3}(2uds-2dus+sdu-dsu+usd-sud)&
\frac{1}{2}(sud+usd-sdu-dsu) \nonumber\\&\Xi^0&\frac{1}{\sqrt
2}(sus-uss)&\frac{1}{\sqrt 6}(sus+uss-2ssu) \nonumber\\
&\Xi^-&\frac{1}{\sqrt 2}(sds-dss)&\frac{1}{\sqrt
6}(sds+dss-2ssd)\nonumber\\\hline (6,1)&\Sigma_{c}^{++}&
\frac{1}{\sqrt 2}(cuu-ucu) &\frac{1}{\sqrt
6}(cuu+ucu-2uuc)\nonumber\\&\Sigma_{c}^{+}&\frac{1}{2}(cud+cdu-ucd-dcu)
&\frac{1}{2\sqrt 3}(dcu+cdu+ucd+cud-2udc-2duc)
\nonumber\\&\Sigma_{c}^{0}&\frac{1}{\sqrt 2}(cdd-dcd) &
\frac{1}{\sqrt 6}(cdd+dcd-2ddc)
\nonumber\\&\Xi_{c}^{'+}&\frac{1}{2}(cus+csu-ucs-scu) &\frac{1}{
2\sqrt 3}(ucs+cus+scu+csu-2usc-2suc) \nonumber\\ &\Xi_{c}^{'0}
&\frac{1}{2}(cds+csd-dcs-scd) &\frac{1}{2\sqrt
3}(dcs+cds+scd+csd-2dsc-2sdc) \nonumber
\\&\Omega_{c}^0&\frac{1}{\sqrt 2}(css-scs) &\frac{1}{\sqrt
6}(scs+css-2ssc)\nonumber \\\hline (\bar 3,
1)&\Lambda_{c}{^+}&\frac{1}{2\sqrt 3}(2udc-2duc+cdu-dcu+ ucd-cud)
&\frac{1}{2}(ucd+cud-dcu-cdu) \nonumber\\
&\Xi_{c}^{+}&\frac{1}{2\sqrt 3}(2usc-2suc+csu-scu+
ucs-cus)&\frac{1}{2}(ucs+cus-scu-csu) \nonumber\\
&\Xi_{c}^{0}&\frac{1}{2\sqrt
3}(2dsc-2sdc+csd-scd+dcs-cds)&\frac{1}{2}(dcs+cds-scd-csd)\nonumber
\\ \hline(3,2)&\Xi_{cc}^{++}&\frac{1}{\sqrt 2}(ucc-cuc)
&\frac{1}{\sqrt 6}(ucc+cuc-2ccu)\nonumber
\\&\Xi_{cc}^{+}&\frac{1}{\sqrt 2}(dcc-cdc)&\frac{1}{\sqrt
6}(dcc+cdc-2ccd)\nonumber \\&\Omega_{cc}^{+}&\frac{1}{\sqrt
2}(scc-csc)&\frac{1}{\sqrt6}(scc+csc-2ccs)\nonumber \\ \hline \ea
\eea}

For the spin $\frac{3}{2}{^+}$ baryons, the flavor wavefunctions
are \bea \ba{ccc}\\\hline & {\rm Baryon} & \varphi^s \nonumber\\
\hline (10,0)&\Delta^{++}&uuu \nonumber \\
&\Delta^{+}&\frac{1}{\sqrt 3}(uud+udu+duu)\nonumber\\
&\Delta^{0}&\frac{1}{\sqrt 3}(udd+ddu+dud)\nonumber\\
&\Delta^{-}&ddd \nonumber\\ &\Sigma^{*+}&\frac{1}{\sqrt
3}(uus+suu+usu)\nonumber\\ &\Sigma^{*-}&\frac{1}{\sqrt
3}(dds+dsd+sdd)\nonumber\\ &\Sigma^{*0}&\frac{1}{\sqrt
6}(sdu+sud+usd+dsu+dus+uds)\nonumber\\ &\Xi^{*0} &\frac{1}{\sqrt
3}(ssu+sus+uss)\nonumber\\ &\Xi^{*-}&\frac{1}{\sqrt
3}(ssd+sds+dss)\nonumber\\ &\Omega^{-}& sss \nonumber \\\hline
(6,1)&\Sigma_{c}^{*++}& \frac{1}{\sqrt 3}(uuc+ucu+cuu)\nonumber\\
&\Sigma_{c}^{*+}& \frac{1}{\sqrt
6}(udc+dcu+cud+cdu+duc+ucd)\nonumber\\ &\Sigma_{c}^{*0}&
\frac{1}{\sqrt 3}(ddc+dcd+cdd)\nonumber\\ &\Xi_{c}^{*+}&
\frac{1}{\sqrt 6}(usc+scu+cus+csu+suc+ucs)\nonumber\\
&\Xi_{c}^{*0}& \frac{1}{\sqrt
6}(dsc+scd+cds+csd+dsc+scd)\nonumber\\ &\Omega_{c}^{*0}&
\frac{1}{\sqrt 3}(ssc+scs+css)\nonumber\\\hline
(3,2)&\Xi_{cc}^{*++}&\frac{1}{\sqrt 3}(ucc+cuc+ccu)\nonumber\\
&\Xi_{cc}^{*+}&\frac{1}{\sqrt 3}(dcc+cdc+ccd)\nonumber\\
&\Omega_{cc}^{*+}& \frac{1}{\sqrt 3}(scc+csc+ccs)\nonumber\\\hline
(1,3)&\Omega_{ccc}^{*++}& ccc\nonumber\\ \hline \ea \eea

We have used the convention $\chi= \chi^{\sigma}_{S_z}$ for the
spin wave functions, where $S_z$ is the third component of the
spin and $\sigma$ represents the symmetry state \be
\chi^{s}_{\frac{3}{2}}=\uparrow\uparrow\uparrow \,,~~~
\chi^{'}_{\frac{1}{2}}=\frac{1}{\sqrt 2}(\uparrow \downarrow
\uparrow- \downarrow \uparrow \uparrow)\,,~~~
\chi^{''}_{\frac{1}{2}}=\frac{1}{\sqrt 6}(2\uparrow \uparrow
\downarrow- \uparrow \downarrow \uparrow- \downarrow \uparrow
\uparrow) \,. \label{chi} \ee Other values of $S_z$ are obtained
by applying the lowering the operator in spin space and
normalizing to unity.

\begin{table}
\begin{center}\tabcolsep -0.1mm \begin{tabular}{|c|c|c|} \hline
Baryon && Valence and sea contribution to the magnetic moments \\
\hline \hline $p$&$\mu_{{\rm val}}$&$\cos^2 \phi
\left[\frac{4}{3}\mu_u- \frac{1}{3}\mu_d\right]+ \sin^2 \phi \left[
\frac{2}{3}\mu_u+ \frac{1}{3}\mu_d \right]$\\ \cline{2-3}
&$\mu_{{\rm sea}}$ &$ \cos^2 \phi \left[- \frac{a}{3}(7+ 4\alpha^2+
\frac{4}{3}\beta^2+ \frac{\zeta^2}{6}+
\frac{17}{4}\gamma^2)\mu_{u}-\frac{a}{3}(2- \alpha^2-
\frac{\beta^2}{3}- \frac{\zeta^2}{24}-
\frac{17}{16}\gamma^2)\mu_{d}- (a \alpha^2)\mu_{s}- (a \gamma^2
)\mu_{c}\right]$\\ &&$+ \sin^2 \phi\left[-\frac{a}{3}(5+ 2\alpha^2+
\frac{2}{3}\beta^2+ \frac{\zeta^2}{12}+
\frac{17}{8}\gamma^2)\mu_{u}- \frac{a}{3}(4+ \alpha^2+
\frac{\beta^2}{3}+ \frac{\zeta^2}{24}+
\frac{17}{16}\gamma^2)\mu_{d}- (a \alpha^2)\mu_{s}-
(a\gamma^2)\mu_{c}\right]$\\ \hline

$\Sigma{^+}$ &$\mu_{{\rm val}}$&$\cos^2 \phi \left[\frac{4}{3}\mu_u
-\frac{1}{3}\mu_s \right]+ \sin^2 \phi \left[\frac{2}{3}\mu_u+
\frac{1}{3}\mu_s \right]$\\ \cline{2-3}&$\mu_{{\rm sea}}$&$ \cos^2
\phi \left[- \frac{a}{3}(8+ 3\alpha^2+ \frac{4}{3} \beta^2 +
\frac{\zeta^2}{6}+ \frac{17}{4} \gamma^2)\mu_{u}- \frac{a}{3}(4-
\alpha^2)\mu_{d}- \frac{a}{3}(2\alpha^2- \frac{4}{3}\beta^2-
\frac{\zeta^2}{24}- \frac{17}{16}\gamma^2)\mu_{s}-
(a\gamma^2)\mu_{c}\right]$\\ &&$+ \sin^2 \phi \left[- \frac{a}{3}(4+
6\alpha^2+ \frac{2}{3}\beta^2+ \frac{\zeta^2}{12}+
\frac{17}{8}\gamma^2)\mu_{u}- \frac{a}{3}(2+ \alpha^2)\mu_{d}-
\frac{a}{3}(4\alpha^2+ \frac{4}{3}\beta^2+ \frac{\zeta^2}{24}+
\frac{17}{16}\gamma^2)\mu_{s}- (a\gamma^2 )\mu_{c}\right]$\\\hline

$\Sigma^0$&$\mu_{{\rm val}}$&$\cos^2 \phi \left[\frac{2}{3}\mu_u+
\frac{2}{3}\mu_d- \frac{1}{3}\mu_s \right]+ \sin^2
\phi\left[\frac{1}{3}\mu_u+ \frac{1}{3}\mu_d+ \frac{1}{3}\mu_s
\right]$\\
\cline{2-3} &$\mu_{{\rm sea}}$& $\cos^2 \phi \left[- \frac{a}{3}(6+
\alpha^2+ \frac{2}{3}\beta^2+ \frac{\zeta^2}{12}
+\frac{17}{8}\gamma^2) \mu_{u}- \frac{a}{3}(6+ \alpha^2+ \frac{
2}{3}\beta^2+ \frac{\zeta^2}{12}+ \frac{17}{8}\gamma^2)\mu_{d}-
\frac{a}{3}(+ 2\alpha^2- \frac{4}{3}\beta^2- \frac{\zeta^2}{24}-
\frac{17}{16}\gamma^2)\mu_{s}- (a\gamma^2) \mu_{c}\right]$\\ &&$+
\sin^2 \phi \left[-\frac{a}{3}(3+ 2\alpha^2+ \frac{\beta^2}{3}+
\frac{\zeta^2}{24}+ \frac{17}{16}\gamma^2 )\mu_{u}-\frac{a}{3}(3+
2\alpha^2+\frac{\beta^2}{3}+ \frac{\zeta^2}{24}+
\frac{17}{16}\gamma^2)\mu_{d}- \frac{a}{3}(4\alpha^2+
\frac{4}{3}\beta^2+ \frac{\zeta^2}{24}+
\frac{17}{16}\gamma^2)\mu_{s}- (a\gamma^2)\mu_{c} \right]$\\\hline

$\Xi^0$&$\mu_{{\rm val}}$&$ \cos^2 \phi \left[-\frac{1}{3}\mu_u+
\frac{4}{3}\mu_s \right]+ \sin^2 \phi \left[\frac{1}{3}\mu_{u}+
\frac{2}{3}\mu_s\right]$\\ \cline{2-3} &$\mu_{{\rm sea}}$& $ \cos^2
\phi \left[- \frac{a}{3}(-2+ 3\alpha^2- \frac{\beta^2}{3}-
\frac{\zeta^2}{24}- \frac{17}{16}\gamma^2)\mu_{u} - \frac{a}{3}(-1+
4\alpha^2)\mu_{d}- \frac{a}{3}(7 \alpha^2+ \frac{16}{3}\beta^2+
\frac{\zeta^2}{6}+ \frac{17}{4}\gamma^2)\mu_{s}- (a\gamma^2)
\mu_{c}\right]$\\ &&$+ \sin^2 \phi \left[- \frac{a}{3}(2+ 3\alpha^2+
\frac{\beta^2}{3} + \frac{\zeta^2}{24}+
\frac{17}{16}\gamma^2)\mu_{u} - \frac{a}{3}(1+ 2\alpha^2)\mu_{d}-
\frac{a}{3}(5\alpha^2+ \frac{8}{3}\beta^2+
\frac{\zeta^2}{12}+\frac{17}{8}\gamma^2)\mu_{s}-(a\gamma^2)
\mu_{c}\right]$\\\hline

$\Lambda$ &$\mu_{{\rm val}}$& $\cos^2 \phi \left[\mu_s\right]+
\sin^2 \phi\left[\frac{1}{3}\mu_{u}+ \frac{1}{3}\mu_{d}+
\frac{1}{3}\mu_s \right]$\\ \cline{2-3} &$\mu_{{\rm sea}}$ &$
\cos^2 \phi \left[-(a \alpha^2)\mu_{u}- (a \alpha^2)\mu_{d}-
a(2\alpha^2+ \frac{4}{3}\beta^2+ \frac{\zeta^2}{24}+
\frac{17}{16}\gamma^2)\mu_{s}- (a \gamma^2)\mu_{c}\right]$\\ &&$
+\sin^2 \phi \left[-\frac{a}{3}(2+ 2\alpha^2+ \frac{\beta^2}{3}+
\frac{\zeta^2}{24}+ \frac{17}{16}\gamma^2)\mu_{u}-\frac{a}{3}(2+
2\alpha^2+ \frac{\beta^2}{3}+ \frac{\zeta^2}{24}+
\frac{17}{16}\gamma^2)\mu_{d}- \frac{a}{3}(4\alpha^2+
\frac{4}{3}\beta^2+ \frac{\zeta^2}{24}+
\frac{17}{16}\gamma^2)\mu_{s}- (a\gamma^2)\mu_{c}\right]$\\\hline

$\Sigma^{++}_{c}$&$\mu_{{\rm val}}$&$\cos^2 \phi
\left[\frac{4}{3}\mu_u - \frac{1}{3}\mu_c\right] + \sin^2 \phi
\left[ \frac{2}{3}\mu_{u}+ \frac{1}{3}\mu_{c} \right]
$\\\cline{2-3}&$\mu_{{\rm sea}}$&$ \cos^2 \phi \left[-
\frac{a}{3}(8+ 4\alpha^2+ \frac{4}{3}\beta^2+ \frac{\zeta^2}{6}+
\frac{13}{4}\gamma^2 )\mu_{u}-\frac{a}{3}(4- \gamma^2)\mu_{d}-
\frac{a}{3}(4\alpha^2-
\gamma^2)\mu_{s}+ \frac{a}{24}(3\zeta^2+ \gamma^2)\mu_{c}\right]$\\
&&$ \sin^2 \phi \left[- \frac{a}{3}(4+ 2\alpha^2+
\frac{2}{3}\beta^2+ \frac{\zeta^2}{12}+
\frac{25}{8}\gamma^2)\mu_{u}- \frac{a}{3}(2+ \gamma^2)\mu_{d}-
\frac{a}{3}(2\alpha^2+ \gamma^2)\mu_{s}- \frac{a}{24}(3\zeta^2+
49\gamma^2)\mu_{c}\right]$\\ \hline

$\Sigma^{+}_{c}$&$\mu_{{\rm val}}$&$\cos^2 \phi
\left[\frac{2}{3}\mu_{u}+ \frac{2}{3}\mu_{d}- \frac{1}{3}\mu_c
\right]+ \sin^2 \phi \left[\frac{1}{3}\mu_{u}+ \frac{1}{3}\mu_{d}+
\frac{1}{3}\mu_{c}\right]$
\\\cline{2-3}&$\mu_{{\rm sea}}$&$\cos^2 \phi \left[- \frac{a}{3}(6+ 2\alpha^2+
\frac{2}{3}\beta^2+ \frac{\zeta^2}{12}+ \frac{9}{8}\gamma^2)\mu_{u}-
\frac{a}{3}(6+ 2\alpha^2+ \frac{2}{3}\beta^2+ \frac{\zeta^2}{12}+
\frac{9}{8}\gamma^2)\mu_{d}- \frac{a}{3}(4\alpha^2-
\gamma^2)\mu_{s}+ \frac{a}{24}(3\zeta^2+ \gamma^2)\mu_{c}\right]$\\
&& $\sin^2 \phi \left[- \frac{a}{3}(3+ \alpha^2+ \frac{\beta^2}{3}+
\frac{\zeta^2}{24}+ \frac{33}{16}\gamma^2)\mu_{u}- \frac{a}{3}(3+
\alpha^2+ \frac{\beta^2 }{3}+ \frac{\zeta^2}{24}+
\frac{33}{16}\gamma^2)\mu_{d}- \frac{a}{3}(2\alpha^2+
\gamma^2)\mu_{s}- \frac{a}{24}(3\zeta^2+ 49\gamma^2)\mu_{c}\right]$\\
\hline

$\Omega^{0}_{c}$&$\mu_{{\rm val}}$&$\cos^2
\phi\left[\frac{4}{3}\mu_{s}- \frac{1}{3}\mu_c \right]+ \sin^2
\phi\left[\frac{2}{3}\mu_{s} + \frac{1}{3}\mu_{c}
\right]$\\\cline{2-3}&$\mu_{{\rm sea}}$&$\cos^2 \phi
\left[-\frac{a}{3}(4\alpha^2- \gamma^2 )\mu_{u}-
\frac{a}{3}(4\alpha^2- \gamma^2 )\mu_{d}- \frac{a}{3}(8\alpha^2+
\frac{16}{3}\beta^2 + \frac{\zeta^2}{6}+
\frac{13}{4}\gamma^2)\mu_{s}+ \frac{a}{24}(3\zeta^2+
\gamma^2)\mu_{c}\right]$\\ &&$\sin^2 \phi \left[-
\frac{a}{3}(2\alpha^2+ \gamma^2)\mu_{u}- \frac{a}{3}(2\alpha^2+
\gamma^2)\mu_{d}- \frac{a}{3}(4\alpha^2+ \frac{8}{3}\beta^2+
\frac{\zeta^2}{12}+ \frac{33}{8}\gamma^2) \mu_{s}-
\frac{a}{24}(3\zeta^2+ 49\gamma^2)\mu_{c}\right]$\\\hline

$\Lambda^{+}_{c}$&$\mu_{{\rm val}}$&$\cos^2 \phi \left[\mu_c\right]+
\frac{1}{3} \sin^2 \phi\left[\mu_{u}+ \mu_{d}+ \mu_{c}\right]$
\\\cline{2-3} &$\mu_{{\rm sea}}$&$\cos^2 \phi \left[-(a\gamma^2)\mu_{u}-
(a\gamma^2)\mu_{d}- (a\gamma^2)\mu_{s}-\frac{3}{8}a(\zeta^2+ 11
\gamma^2 )\mu_{c}\right]$\\&&$+ \sin^2 \phi \left[-\frac{a}{3}(3+
\alpha^2 + \frac{\beta^2}{3}+ \frac{\zeta^2}{24}+
\frac{33}{16}\gamma^2)\mu_{u}- \frac{a}{3}(3+ \alpha^2+
\frac{\beta^2}{3}+ \frac{\zeta^2}{24}+
\frac{33}{16}\gamma^2)\mu_{d}- \frac{a}{3}(2 \alpha^2+
\gamma^2)\mu_{s}- \frac{a}{24}(3\zeta^2+ 49\gamma^2)\mu_{c}\right]$\\
\hline

$\Xi^{+}_{c}$&$\mu_{{\rm val}}$&$ \cos^2 \phi [\mu_c]+
\frac{1}{3}\sin^2 \phi \left[\mu_{u}+ \mu_{s}+ \mu_{c}\right]$
\\\cline{2-3} &$\mu_{{\rm sea}}$&$ \cos^2 \phi\left[-(a\gamma^2)\mu_{u}-
(a\gamma^2)\mu_{d}-
(a\gamma^2)\mu_{s}- \frac{3}{8}a(\zeta^2+ 11\gamma^2)\mu_{c}\right]$\\
&&$+ \sin^2 \phi \left[-\frac{a}{3}(2+ 2\alpha^2+ \frac{\beta^2}{3}+
\frac{\zeta^2}{24}+ \frac{33}{16}\gamma^2)\mu_{u}- \frac{a}{3}(1+
\alpha^2+ \gamma^2)\mu_{d}- \frac{a}{3}(3\alpha^2+
\frac{4}{3}\beta^2+ \frac{\zeta^2}{24}+
\frac{33}{16}\gamma^2)\mu_{s}- \frac{a}{24}(3\zeta^2+
49\gamma^2)\mu_{c}\right]$\\\hline

$\Omega^{+}_{cc}$&$\mu_{{\rm val}}$&$\cos^2 \phi \left[-
\frac{1}{3}\mu_s+ \frac{4}{3}\mu_c\right] +\sin^2
\phi\left[\frac{1}{3}\mu_{s}+ \frac{2}{3}\mu_{c} \right]$
\\\cline{2-3}&$\mu_{{\rm sea}}$ &$\cos^2 \phi \left[\frac{a}{3}(\alpha^2- 4
\gamma^2)\mu_{u}+ \frac{a}{3}(\alpha^2- 4\gamma^2)\mu_{d}+
\frac{a}{3}(2\alpha^2+ \frac{4}{3}\beta^2+ \frac{\zeta^2}{24}-
\frac{47}{16}\gamma^2)\mu_{s}- \frac{a}{2}(\zeta^2+
\frac{31}{3}\gamma^2)\mu_{c}\right]$\\&&$+ \sin^2 \phi \left[ -
\frac{a}{3}(\alpha^2+ 2\gamma^2)\mu_{u}- \frac{a}{3}(\alpha^2 +
2\gamma^2)\mu_{d}- \frac{a}{3}(2\alpha^2+ \frac{4}{3}\beta^2+
\frac{\zeta^2}{24}+ \frac{49}{16}\gamma^2)\mu_{s}-
\frac{a}{4}(\zeta^2+ \frac{37}{3}\gamma^2)\mu_{c}\right]$\\\hline
\end{tabular}
\caption{Valence and sea contribution of the charmed spin
$\frac{1}{2}{^+}$ baryons in terms of $\chi$CQM parameters and
configuration mixing parameter $\phi$. The spin polarizations for
the other baryons can be found from isospin symmetry.}
\label{spin1/2charm}
\end{center}
\end{table}

\begin{table}
\begin{center}
\begin{tabular}{|c|c|c|}  \hline

Baryon & & Valence and sea contribution to the magnetic moments \\
\hline \hline $\Delta^{++}$ &$\mu_{{\rm val}}$&3$\mu_{u}$\\
\cline{2-3} &$\mu_{{\rm sea}}$&$-a \left(6+ 3\alpha^2+ \beta^2+
\frac{\zeta^2}{8}+ \frac{51}{16}\gamma^2\right)\mu_{u}- 3a\mu_{d}
-3a\alpha^2 \mu_{s}-3a \gamma^2\mu_{c}$\\\hline

$\Delta^{+}$&$\mu_{{\rm val}}$& $2\mu_{u}+ \mu_{d}$\\\cline{2-3}
&$\mu_{{\rm sea}}$&$-a \left(5 + 2\alpha^2+ \frac{2}{3}\beta^2+
\frac{\zeta^2}{12}+ \frac{17}{8}\gamma^2\right)\mu_{u}- a\left(4+
\alpha^2+ \frac{\beta^2}{3}+ \frac{\zeta^2}{24}+
\frac{17}{16}\gamma^2\right)\mu_{d}- 3a\alpha^2\mu_{s}- 3a\gamma^2
\mu_{c}$\\\hline

$\Sigma^{*+}$&$\mu_{{\rm val}}$& $2\mu_{u}+ \mu_{s}$\\\cline{2-3}
&$\mu_{{\rm sea}}$&$-a \left(4+ 3\alpha^2+ \frac{2}{3}\beta^2+
\frac{\zeta^2}{12}+ \frac{17}{8}\gamma^2\right)\mu_{u}- a\left(2 +
\alpha^2 \right)\mu_{d}- a\left(4\alpha^2+ \frac{4}{3}\beta^2+
\frac{\zeta^2}{24}+ \frac{17}{16}\gamma^2\right) \mu_{s}-
3a\gamma^2\mu_{c}$\\\hline

$\Sigma^{*0}$&$\mu_{{\rm val}}$& $\mu_{u}+ \mu_{d}+
\mu_{s}$\\\cline{2-3}&$\mu_{{\rm sea}}$&$-a \left(3+2\alpha^2+
\frac{\beta^2}{3}+ \frac{\zeta^2}{24}+
\frac{17}{16}\gamma^2\right)\mu_{u}- a\left(3+ 2\alpha^2+
\frac{\beta^2}{3}+ \frac{\zeta^2}{24}+ \frac{17}{16}\gamma^2
\right)\mu_{d}- a\left(4\alpha^2+ \frac{4}{3}\beta^2+
\frac{\zeta^2}{24}+ \frac{17}{16}\gamma^2\right)\mu_{s} -3a\gamma^2
\mu_{c}$\\\hline

$\Xi^{*0}$&$\mu_{{\rm val}}$&$ \mu_{u}+ 2\mu_{s}$\\\cline{2-3}
&$\mu_{{\rm sea}}$&$-a \left(2+ 3\alpha^2+ \frac{\beta^2}{3}+
\frac{\zeta^2}{24}+ \frac{17}{16}\gamma^2\right)\mu_{u}- a\left(1+
2\alpha^2\right) \mu_{d}- a\left(5\alpha^2+ \frac{8}{3}\beta^2+
\frac{\zeta^2}{12}+ \frac{17}{8}\gamma^2\right)\mu_{s}-
3a\gamma^2\mu_{c}$\\\hline

$\Omega^{-}$&$\mu_{{\rm val}}$&3$\mu_{s}$
\\\cline{2-3}&$\mu_{{\rm sea}}$&$-3a\alpha^2\mu_{u} -3a\alpha^2\mu_{d}-
a\left(6\alpha^2+ 4\beta^2+ \frac{\zeta^2}{8} +
\frac{51}{16}\gamma^2\right)\mu_{s}- 3a\gamma^2\mu_{c}$\\\hline

$\Sigma^{*++}_{c}$ &$\mu_{{\rm val}}$&$2\mu_{u}+
\mu_{c}$\\\cline{2-3} &$\mu_{{\rm sea}}$&$-a\left(4+ 2\alpha^2+
\frac{2}{3}\beta^2+ \frac{\zeta^2}{12}+
\frac{25}{8}\gamma^2\right)\mu_{u}- a\left(2+ \gamma^2
\right)\mu_{d}- a\left(2\alpha^2+ \gamma^2 \right)\mu_{s}-
\frac{a}{8}\left(3\zeta^2+ 49\gamma^2 \right)\mu_{c}$\\ \hline

$\Sigma^{*+}_{c}$ &$\mu_{{\rm val}}$&$\mu_{u}+ \mu_{d}+
\mu_{c}$\\\cline{2-3} &$\mu_{{\rm sea}}$& $-a\left(3+ \alpha^2+
\frac{\beta^2}{3}+ \frac{\zeta^2}{24}+
\frac{33}{16}\gamma^2\right)\mu_{u}- a\left(3+ \alpha^2+
\frac{\beta^2}{3}+ \frac{\zeta^2}{24}+ \frac{33}{16}
\gamma^2\right)\mu_{d}- a\left(2 \alpha^2+ \gamma^2\right)
\mu_{s}- \frac{a}{8}\left(3\zeta^2+ 49\gamma^2\right)\mu_{c}$\\
\hline

$\Xi^{*+}_{c}$ &$\mu_{{\rm val}}$&$\mu_{u}+ \mu_{s}+
\mu_{c}$\\\cline{2-3} &$\mu_{{\rm sea}}$&$-a\left(2+ 2\alpha^2+
\frac{\beta^2}{3}+ \frac{\zeta^2}{24}+
\frac{33}{16}\gamma^2\right)\mu_{u}- a\left(1+ \alpha^2
+\gamma^2\right) \mu_{d}- a\left (3\alpha^2+ \frac{4}{3}\beta^2
+\frac{\zeta^2}{24}+ \frac{33}{16}\gamma^2\right)\mu_{s} -
\frac{a}{8}\left(3\zeta^2+ 49\gamma^2\right)\mu_{c}$\\ \hline

$\Omega^{*0}_{c}$ &$\mu_{{\rm val}}$&$2\mu_{s}+
\mu_{c}$\\\cline{2-3}&$\mu_{{\rm sea}}$& $- a(2\alpha^2+
\gamma^2)\mu_{u}- a(2\alpha^2+ \gamma^2)\mu_{d}- a(4\alpha^2+
\frac{8}{3}\beta^2+ \frac{\zeta^2}{12}+
\frac{25}{8}\gamma^2)\mu_{s}- \frac{a}{8}\left(3\zeta^2+ 49\gamma^2
\right)\mu_{c}$\\ \hline

$\Xi^{*++}_{cc}$& $\mu_{{\rm val}}$&$\mu_{u}+ 2\mu_{c}$\\\cline{2-3}
&$\mu_{{\rm sea}}$&$-a\left(2+ \alpha^2+ \frac{\beta^2}{3}+
\frac{\zeta^2}{24}+ \frac{49}{16}\gamma^2\right)\mu_{u}- a\left(1+
2\gamma^2\right)\mu_{d}- a \left(\alpha^2+ 2\gamma^2 \right)\mu_{s}
- \frac{a}{4}\left(3\zeta^2+ 37\gamma^2\right)\mu_{c}$\\\hline

$\Omega^{*+}_{cc}$&$\mu_{{\rm val}}$&$\mu_{s}+
2\mu_{c}$\\\cline{2-3} &$\mu_{{\rm sea}}$& $- a\left(\alpha^2+
2\gamma^2\right)\mu_{u}- a\left(\alpha^2+ 2\gamma^2\right)\mu_{d}-
a\left(2\alpha^2+ \frac{4}{3} \beta^2+\frac{\zeta^2}{24}+
\frac{49}{16}\gamma^2\right)\mu_{s} - \frac{a}{4}\left(3\zeta^2+
37\gamma^2\right)\mu_{c}$\\\hline

$\Omega^{*++}_{ccc}$&$\mu_{{\rm val}}$&$3\mu_{c}$\\
\cline{2-3}&$\mu_{{\rm sea}}$& $-3a\gamma^2 \mu_{u}-3a\gamma^2
\mu_{d}- 3a\gamma^2\mu_{s}- \frac{9}{8}a\left(\zeta^2+
11\gamma^2\right)\mu_{c}$\\\hline
\end{tabular}\caption{Valence and sea contributions of the low
lying and charmed spin $\frac{3}{2}{^+}$ baryons in terms of the
$\chi$CQM parameters. The spin polarizations for the other baryons
can be found from isospin symmetry.}\label{spin3/2}
\end{center}
\end{table}

\footnotesize{
\begin{sidewaystable}
\begin{center}
\begin{tabular}{|c|c|c|} \hline
Transition & & Valence and sea contribution to the transition
magnetic moments \\ \hline \hline

$\Delta p$&$\mu_{{\rm val}}$&$\frac{2\sqrt{2}}{3}\mu_{u} -
\frac{2\sqrt{2}}{3}\mu_{d}$\\\cline{2-3} &$\mu_{{\rm sea}}$& $-
\frac{2\sqrt{2}}{3}a\left(1+ \alpha^2+ \frac{\beta^2}{3}+
\frac{\zeta^2}{24}+ \frac{17}{16}\gamma^2\right)\mu_{u}+
\frac{2\sqrt{2}}{3}a\left(1+ \alpha^2+ \frac{\beta^2}{3}+
\frac{\zeta^2}{24}+ \frac{17}{16}\gamma^2\right)\mu_{d}$ \\\hline

$\Sigma^{*+} \Sigma^{+}$&$\mu_{{\rm val}}$&
$\frac{2\sqrt{2}}{3}\mu_{u}-
\frac{2\sqrt{2}}{3}\mu_{s}$\\\cline{2-3} &$\mu_{{\rm
sea}}$&$-\frac{2\sqrt{2}}{3}a\left(2+ \frac{\beta^2}{3}+
\frac{\zeta^2}{24}+ \frac{17}{16}\gamma^2\right)\mu_{u}
-\frac{2\sqrt{2}}{3}a\left(1- \alpha^2\right)\mu_{d}+
\frac{2\sqrt{2}}{3}a \left(\alpha^2+ \frac{4}{3}\beta^2+
\frac{\zeta^2}{24}+ \frac{17}{16}\gamma^2\right)\mu_{s}$\\ \hline

$\Sigma^{*0} \Sigma^{0}$&$\mu_{{\rm
val}}$&$\frac{\sqrt{2}}{3}\mu_{u}+ \frac{\sqrt{2}}{3}\mu_{d}-
\frac{2\sqrt{2}}{3}\mu_{s}$\\\cline{2-3} &$\mu_{{\rm
sea}}$&$-\frac{\sqrt{2}}{3}a\left(3- \alpha^2+ \frac{\beta^2}{3}+
\frac{\zeta^2}{24}+ \frac{17}{16}\gamma^2\right) \mu_{u}
-\frac{\sqrt{2}}{3}a\left(3- \alpha^2+ \frac{\beta^2}{3}+
\frac{\zeta^2}{24}+ \frac{17}{16} \gamma^2\right)\mu_{d}+
\frac{2\sqrt{2}}{3}a\left(\alpha^2+ \frac{4}{3}\beta^2+
\frac{\zeta^2}{24}+ \frac{17}{16}\gamma^2\right)\mu_{s}$\\ \hline

$\Xi^{*0} \Xi^{0}$&$\mu_{{\rm val}}$&$\frac{2\sqrt{2}}{3}\mu_{u}-
\frac{2\sqrt{2}}{3}\mu_{s}$\\\cline{2-3}&$\mu_{{\rm sea}}$&
$-\frac{2\sqrt{2}}{3}a\left(2+ \frac{\beta^2}{3}+ \frac{\zeta^2}{24}
+ \frac{17}{16}\gamma^2\right)\mu_{u}- \frac{2\sqrt{2}}{3}a\left(1-
\alpha^2 \right)\mu_{d}+ \frac{2\sqrt{2}}{3}a\left(\alpha^2+
\frac{4}{3}\beta^2+ \frac{\zeta^2}{24}+ \frac{17}{16}
\gamma^2\right)\mu_{s}$\\ \hline

$\Sigma^{*0} \Lambda$&$\mu_{{\rm val}}$&$\sqrt{\frac{2}{3}}\mu_{u}-
\sqrt{\frac{2}{3}}\mu_{d}$\\\cline{2-3}&$\mu_{{\rm
sea}}$&$-\sqrt{\frac{2}{3}} a\left(1+ \alpha^2+ \frac{\beta^2}{3}+
\frac{\zeta^2}{24} + \frac{17}{16}\gamma^2\right)\mu_{u}+
\sqrt{\frac{2}{3}}a\left(1+ \alpha^2+ \frac{\beta^2}{3}+
\frac{\zeta^2}{24}+ \frac{17}{16} \gamma^2\right)\mu_{d}$\\\hline

$\Sigma^{*++}_{c} \Sigma^{++}_{c} $&$\mu_{{\rm
val}}$&$-\frac{2\sqrt{2}}{3}
\mu_{u}+\frac{2\sqrt{2}}{3}\mu_{c}$\\\cline{2-3} &$\mu_{{\rm
sea}}$&$ \frac{2\sqrt{2}}{3}a\left(2+ \alpha^2+ \frac{\beta^2}{3}+
\frac{\zeta^2}{24}+ \frac{\gamma^2}{16}\right)\mu_{u}+
\frac{2\sqrt{2}}{3}a\left(1- \gamma^2\right)\mu_{d}+
\frac{2\sqrt{2}}{3}a\left(\alpha^2- \gamma^2\right)\mu_{s}-
\frac{\sqrt{2}}{12}a\left(3\zeta^2+ 25\gamma^2\right)\mu_{c}$
\\\hline

$\Sigma^{*+}_{c} \Sigma^{+}_{c}$&$\mu_{{\rm val}}$&$-
\frac{\sqrt{2}}{3}\mu_{u}- \frac{\sqrt{2}}{3}\mu_{d}+
\frac{2\sqrt{2}}{3}\mu_{c}$\\\cline{2-3} &$\mu_{{\rm sea}}$&
$\frac{\sqrt{2}}{3}a\left(2+ \alpha^2+ \frac{\beta^2}{3}+
\frac{\zeta^2}{24}- \frac{15}{16}\gamma^2\right)\mu_{u}+
\frac{\sqrt{2}}{3} a\left(2+ \alpha^2+ \frac{\beta^2}{3}+
\frac{\zeta^2}{24}- \frac{15}{16}\gamma^2\right)\mu_{d}+
\frac{2\sqrt{2}}{3}a\left(\alpha^2- \gamma^2\right)\mu_{s}-
\frac{\sqrt{2}}{12}a\left(3\zeta^2+ 25 \gamma^2\right)\mu_{c}$
\\\hline

$\Omega^{*0}_{c} \Omega^{0}_{c}$&$\mu_{{\rm val}}$&$
-\frac{2\sqrt{2}}{3}\mu_{s}+
\frac{2\sqrt{2}}{3}\mu_{c}$\\\cline{2-3} &$\mu_{{\rm sea}}$&
$\frac{2\sqrt{2}}{3}a\left(\alpha^2- \gamma^2\right)\mu_{u}+
\frac{2\sqrt{2}}{3}a\left(\alpha^2- \gamma^2\right)\mu_{d}+
\frac{2\sqrt{2}}{3}a\left(2\alpha^2 + \frac{4}{3}\beta^2+
\frac{\zeta^2}{24}+ \frac{\gamma^2}{16}\right)\mu_{s}-
\frac{\sqrt{2}}{12}a\left(3\zeta^2+ 25 \gamma^2\right)\mu_{c}$
\\\hline

$\Sigma^{*+}_{c} \Lambda^{+}_{c}$&$\mu_{{\rm
val}}$&$\sqrt{\frac{2}{3}}\mu_{u}-
\sqrt{\frac{2}{3}}\mu_{d}$\\\cline{2-3}&$\mu_{{\rm sea}}$&
$-\sqrt{\frac{2}{3}}a\left(1+ \alpha^2+ \frac{\beta^2}{3}+
\frac{\zeta^2}{24}+ \frac{17}{16}\gamma^2\right)\mu_{u}+
\sqrt{\frac{2}{3}}a\left(1+ \alpha^2+ \frac{\beta^2}{3}+
\frac{\zeta^2}{24}+ \frac{17}{16}\gamma^2\right) \mu_{d}$\\ \hline

$\Xi^{*+}_{c} \Xi^{+}_{c}$&$\mu_{{\rm
val}}$&$\sqrt{\frac{2}{3}}\mu_{u}- \sqrt{\frac{2}{3}}\mu_{s}$
\\\cline{2-3} &$\mu_{{\rm sea}}$&$ -\sqrt{\frac{2}{3}} a\left(2+
\frac{\beta^2}{3}+ \frac{\zeta^2}{24}+ \frac{17}{16}
\gamma^2\right)\mu_{u}- \sqrt{\frac{2}{3}}a\left( 1-
\alpha^2\right)\mu_{d}+ \sqrt{\frac{2}{3}}a\left( \alpha^2+
\frac{4}{3}\beta^2+ \frac{\zeta^2}{24}+ \frac{17}{16}\gamma^2
\right)\mu_{s} $\\\hline

$\Xi^{*++}_{cc} \Xi^{++}_{cc}$&$\mu_{{\rm
val}}$&$\frac{2\sqrt{2}}{3}\mu_{u}-
\frac{2\sqrt{2}}{3}\mu_{c}$\\\cline{2-3}&$\mu_{{\rm sea}}$&$
-\frac{2\sqrt{2}}{3}a\left(2+ \alpha^2+ \frac{\beta^2}{3} +
\frac{\zeta^2}{24}+ \frac{\gamma^2}{16}\right)\mu_{u}-
\frac{2\sqrt{2}}{3}a\left(1- \gamma^2\right)\mu_{d}-
\frac{2\sqrt{2}}{3}a\left(\alpha^2- \gamma^2\right)\mu_{s}+
\frac{\sqrt{2}}{12}a\left(3\zeta^2+ 25 \gamma^2\right)\mu_{c}$
\\\hline

$\Omega^{*+}_{cc} \Omega^{+}_{cc}$&$\mu_{{\rm
val}}$&$\frac{2\sqrt{2}}{3}\mu_{s}-
\frac{2\sqrt{2}}{3}\mu_{c}$\\\cline{2-3}&$\mu_{{\rm sea}}$&$
-\frac{2\sqrt{2}}{3}a\left(\alpha^2- \gamma^2\right)\mu_{u}-
\frac{2\sqrt{2}}{3}a\left(\alpha^2- \gamma^2\right)\mu_{d} -
\frac{2\sqrt{2}}{3}a\left(2\alpha^2+ \frac{4}{3}\beta^2 +
\frac{\zeta^2}{24}+ \frac{\gamma^2}{16}\right)\mu_{s}+
\frac{\sqrt{2}}{12}a\left(3\zeta^2+ 25\gamma^2\right)$
\\\hline\hline

$\Sigma^{0} \Lambda$&$\mu_{{\rm val}}$&$\frac{1}{\sqrt
3}\mu_{u}-\frac{1}{\sqrt 3}\mu_{d}$\\\cline{2-3} &$\mu_{{\rm
sea}}$&$-\frac{a}{\sqrt{3}}\left( 1+ \alpha^2+ \frac{\beta^2}{3}+
\frac{\zeta^2}{24}+ \frac{17}{16}\gamma^2 \right)\mu_{u}+
\frac{a}{\sqrt{3}} \left(1+ \alpha^2+ \frac{\beta^2}{3}+
\frac{\zeta^2}{24}+ \frac{17}{16}\gamma^2\right) \mu_{d}$\\ \hline

$\Lambda^{+}_{c} \Sigma^{+}_{c}$&$\mu_{{\rm
val}}$&$\frac{1}{\sqrt{3}}\mu_{u} -
\frac{1}{\sqrt{3}}\mu_{d}$\\\cline{2-3} &$\mu_{{\rm sea}}$&$ -
\frac{1}{\sqrt{3}}a\left(1+ \alpha^2+ \frac{\beta^2}{3}+
\frac{\zeta^2}{24}+ \frac{17}{16}\gamma^2\right)\mu_{u} +
\frac{1}{\sqrt{3}}a\left(1+ \alpha^2+ \frac{\beta^2}{3}+
\frac{\zeta^2}{24}+ \frac{17}{16}\gamma^2\right) \mu_{d}$\\ \hline

$\Xi^{'+}_{c} \Xi^{+}_{c} $&$\mu_{{\rm val}}$
&$\frac{1}{\sqrt{3}}\mu_{u}-
\frac{1}{\sqrt{3}}\mu_{s}$\\\cline{2-3}&$\mu_{{\rm sea}}$&
$-\frac{1}{\sqrt{3}}a\left(2+ \frac{\beta^2}{3}+
\frac{\zeta^2}{24}+ \frac{17}{16} \gamma^2\right)\mu_{u}-
\frac{1}{\sqrt{3}}a\left(1- \alpha^2\right)\mu_{d}+
\frac{1}{\sqrt{3}}a\left(\alpha^2+ \frac{4}{3}\beta^2+
\frac{\zeta^2}{24}+ \frac{17}{16}\gamma^2 \right)\mu_{s}$\\ \hline
\end{tabular}
\caption{ Valence and sea contributions of the low lying  and
charmed spin $\frac{3}{2}{^+} \to \frac{1}{2}{^+}$ and spin
$\frac{1}{2}{^+} \to \frac{1}{2}{^+}$ transition magnetic moments
in terms of the $\chi$CQM parameters. The spin polarizations for
the other transitions can be found from isospin
symmetry.}\label{trans}
\end{center}
\end{sidewaystable}}

\begin{table}
\begin{center}
\begin{tabular}{|c|c|c|c|c|c|c|c|c|c|}\hline
Baryon & Data & NRQM & Lattice QCD &QCDSR \cite{wanglee}& LCQSR &
Valence & Sea & Orbital & Total\\ & \cite{pdg}&\cite{choudhury}&
\cite{lattice}& QSSR \cite{qssr}& \cite{lcqsr} &&&& \\\hline
\hline $\mu(p)$&2.79$\pm 0.00$&3&2.793 &2.82 $\pm$ 0.26&2.7
$\pm$0.5&2.90&$-$0.58&0.48 &2.80\\ $\mu(n)$&$-$1.91$\pm$0.00&$-$2
&$-1.59\pm 0.21$&$-$1.97 $\pm$ 0.15&$-1.8\pm$0.35 &
$-$1.85&0.18&$-$0.44&$-$2.11\\
$\mu(\Sigma^{+})$&2.46$\pm$0.01&2.88& 2.37$\pm$0.18 &2.31
$\pm$0.25&2.2$\pm$0.4&2.50&$-$0.51&0.40&2.39\\
$\mu(\Sigma^{0})$&... &0.88&0.65 $\pm$ 0.06& 0.69 $\pm$0.07&
0.5$\pm$ 0.10&0.74&$-$0.22&0.02& 0.54\\ $\mu(\Sigma^{-})$&$-$1.16
$\pm$0.025& $-$1.12&$-1.07 \pm 0.11$&$-$1.16 $\pm$0.10& $-0.8\pm
0.2$& $-$1.02&0.06&$-$0.36&$-$1.32\\ $\mu(\Xi^{0})$& $-$1.25
$\pm$0.014 &$-$ 1.53& $-1.17 \pm 0.10$& $-$1.17$\pm$ 0.10&$-1.3
\pm$0.3 &$-$1.29&0.14&$-$0.09&$-$1.24\\
$\mu(\Xi^{-})$&$-$0.651$\pm$0.003& $-$0.53&$- 0.51 \pm
0.07$&$-$0.64$\pm$ 0.06& $-0.7 \pm 0.2$&$-$0.59&0.03&0.06&
$-$0.50\\ $\Delta$CG&0.49$\pm$ 0.05 &0.0&...&...&...&0.53
&$-$0.08&0.01&0.46\\
$\mu(\Lambda)$&$- 0.613 \pm$0.004&$-$0.65 &$-0.50 \pm 0.07$&$- 0.56\pm$0.15&$-0.7\pm
0.2 $&$-$0.59&0.02&$-$0.01& $-$0.58\\\hline
$\mu(\Sigma_{c}^{++})$&...& 2.54&...&2.1$\pm$ 0.3
&...&2.32&$-$0.52&0.40&2.20 \\ $\mu(\Sigma_{c}^{+})$&...& 0.54&...&0.6
$\pm$0.1 &...&0.51&$-$0.23&0.02&0.30\\ $\mu(\Sigma_{c}^{0})$&...&
$-$1.46&...&$-1.6 \pm 0.2$&...& $-$1.30&0.06&$-$0.36&$-$1.60\\
$\mu(\Xi_{c}^{'+})$&...& 0.77&...&...&...& 0.78&$-$0.21&0.19&0.76 \\
$\mu(\Xi_{c}^{'0})$&...& $-$1.23&...&...&...&
$-$1.16&0.03&$-$0.19&$-$1.32\\ $\mu(\Omega_{c}^{0})$&...&
$-$0.99&...&...&...&$-$0.93&0.04&$-$0.01&$-$0.90 \\
$\mu(\Lambda_{c}^{+})$&...& 0.39&...&0.15$\pm$ 0.05& 0.40$\pm$ 0.05 &
0.409&$-$0.019&0.002&0.392 \\ $\mu(\Xi_{c}^{+})$&...&0.39&...&...&0.50
$\pm$0.05 &0.41&$-$0.02& 0.01&0.40  \\ $\mu(\Xi_{c}^{0})$&...&
0.39&...&...&0.35$\pm$ 0.05&0.29&$-$0.0003&$-$0.01&0.28\\ \hline
$\mu(\Xi_{cc}^{++})$&...&$-$0.15&...&...&...&$-$0.025&0.111&$-$0.080&0.006\\
$\mu(\Xi_{cc}^{+})$&...&0.85&...&...&...&0.79&$-$0.02&0.07&0.84\\
$\mu(\Omega_{cc}^{+})$&...& 0.73&...&...&...& 0.706&$-$0.013&0.004&0.697\\
\hline
\end{tabular}
\caption{Magnetic moment of the low lying and charmed spin
$\frac{1}{2}{^+}$ baryons with configuration mixing (in units of
$\mu_{N}$).} \label{spin1/2num}
\end{center}
\end{table}

\begin{table}
\begin{center}
\begin{tabular}{|c|c|c|c|c|c|c|c|c|c|}\hline
Baryon & Data & NRQM &Lattice QCD & QCDSR & LCQSR & Valence & Sea
& Orbital & Total\\ & \cite{pdg} &\cite{choudhury} &
\cite{lattice} & \cite{wanglee} & \cite{tam3/2} &&&&
\\\hline\hline $\mu(\Delta^{++})$&3.7 $\sim$ 7.5& 6&4.99$\pm
0.56$& 4.13$\pm$1.30&4.4$\pm$ 0.8&4.53&$-$0.97&0.95&4.51\\
$\mu(\Delta^{+})$& $2.7^{+1.0}_{-1.3} \pm 1.5\pm 3$
\cite{kotulla}& 3&2.49$\pm$0.27& 2.07$\pm$0.65&
2.2$\pm$0.4&2.27&$-$0.61&0.34&2.00\\ $\mu(\Delta^{0})$&...&0.0
&0.06$\pm$ 0.0&0.0&0.0& 0.0& $-$0.25&$-$0.26&$-$0.51\\
$\mu(\Delta^{-})$&...&$-$3&$-$2.45$\pm$0.27&$-
2.07\pm$0.65&$-$2.2$\pm$0.4& $-$2.27& 0.12& $-$0.87&$-3.02$\\
$\mu(\Sigma^{*+})$&...&3.35&2.55$\pm$ 0.26&2.13$\pm$0.82 &
2.7$\pm$0.6&2.74&$-$0.67&0.62&2.69\\ $\mu(\Sigma^{*0})$&...&0.35&
0.27$\pm$0.05&$0.32\pm0.15$&0.20$\pm$0.05& 0.29& $-$0.29&
0.02&0.02\\ $\mu(\Sigma^{*-})$&...& $-$2.65& $-2.02 \pm
0.18$&$-1.66\pm0.73$ &$-$2.28$\pm$0.5&$-$2.16& 0.11&
$-$0.59&$-$2.64\\ $\mu(\Xi^{*0})$&...&0.71&0.46$\pm$0.07&$0.69\pm
0.29$&0.40$\pm$0.08& 0.51& $-$0.26& 0.29&0.54\\
$\mu(\Xi^{*-})$&...&$-$2.29 & $-1.68\pm 0.12$&$-1.51 \pm 0.52$ &
$-$2.0$\pm$0.4&$-$1.64&0.08& $-$0.31&$-$1.87\\
$\mu(\Omega^{-})$&$-$2.02 $\pm$0.06 &$-$1.94& $-1.40\pm 0.10$&
$-1.49 \pm 0.45$&$-$1.65$\pm$0.35& $-$1.76& 0.08&$-$0.03&
$-$1.71\\ & $-$1.94 $\pm$ 0.31 \cite{diehl}&&&&&&&&\\\hline
$\mu(\Sigma_{c}^{*++})$&...& 4.39&...&...&$4.81\pm1.22$
&4.09&$-$0.80&0.63&3.92\\ $\mu(\Sigma_{c}^{*+})$&...&
1.39&...&...&$2.00\pm0.46$ &1.30&$-$0.36&0.03&0.97\\
$\mu(\Sigma_{c}^{*0})$&...&$-$1.61&...&...&$-0.81\pm0.20$&
$-$1.50&0.09&$-$0.58&$-$1.99\\
$\mu(\Xi_{c}^{*+})$&...&1.74&...&...&$1.68\pm0.42$&
1.67&$-$0.39&0.31 &1.59\\
$\mu(\Xi_{c}^{*0})$&...&$-$1.26&...&...&$-0.68\pm0.18$& $-$1.21&
0.08& $-$0.30&$-$ 1.43\\ $\mu(\Omega_{c}^{*0})$&...&
$-$0.91&...&...&$-0.62\pm0.18$& $-$0.89&
0.05&$-$0.02&$-$0.86\\\hline
$\mu(\Xi_{cc}^{*++})$&...&2.78&...&...&...&2.78&$-$0.44&
0.32&2.66\\ $\mu(\Xi_{cc}^{*+})$&...&$-$0.22&...&...&...&$-$
0.22&0.04&$-$0.29&$-$0.47\\ $\mu(\Omega_{cc}^{*+})$&...&
0.13&...&...&...&0.13&0.02&$-$0.01&0.14\\ \hline
$\mu(\Omega_{ccc}^{*++})$&...&
1.17&...&...&...&0.165&0.011&$-$0.002&0.155\\ \hline
\end{tabular}
\caption{The magnetic moments of the low lying and charmed spin
$\frac{3}{2}{^+}$ baryons (in units of $\mu_{N}$).}
\label{spin3/2num}
\end{center}
\end{table}

\begin{table}\begin{center}\begin{tabular}{|c|c|c|c|c|c|c|c|c|c|} \hline
\multicolumn{10}{|c|}{Spin $\frac{3}{2}{^+} \to \frac{1}{2}{^+}$
transitions} \\\hline \hline & Transition & Data & NRQM
\cite{choudhury} & Lattice QCD \cite{lattice} & LCQSR \cite{tam} &
Valence & Sea & Orbital & Total\\ \hline 10 $\to$ 8& $\mu(\Delta
p)$ & 3.46$\pm$ 0.03 \cite{tiator}& 2.65&2.46 $\pm$ 0.43&2.5
$\pm$1.3&2.78&$-$0.44&0.53 &2.87\\ &$\mu(\Sigma^{*+}
\Sigma^{+})$&...&2.42& 2.61 $\pm$ 0.35
&2.1$\pm$0.85&2.38&$-$0.41&0.29&2.26\\ &$\mu(\Sigma^{*0}
\Sigma^{0})$&...&1.05& 1.07 $\pm$ 0.13&0.89
$\pm$0.38&1.03&$-$0.20&0.02&0.85\\ &$\mu(\Sigma^{*-}
\Sigma^{-})$&...&$-$0.32& $- 0.47 \pm 0.09$&$-$0.31 $\pm$
0.10&$-$0.32&0.02&$-$0.25&$-$0.55\\ &$\mu(\Xi^{*0}
\Xi^{0})$&...&2.18& $- 2.77 \pm 0.31$&2.2 $\pm$
0.74&2.24&$-$0.39&0.27&2.12\\ &$\mu(\Xi^{*-} \Xi^{-})$
&...&$-$0.29& 0.47 $\pm$ 0.08&$-$0.31
$\pm$0.11&$-$0.26&0.02&$-$0.23&$-$0.47\\ &$\mu(\Sigma^{*0}
\Lambda)$&...&2.31&...&2.3 $\pm$1.4&2.42&$-$0.39&0.47&2.50
\\\hline 6$\to$6&$\mu(\Sigma^{*++}_{c}
\Sigma^{++}_{c})$&...&$-$1.51&...&$-$2.8$\pm$1.0&$-$1.45&0.38&$-$0.30&$-$1.37\\
&$\mu(\Sigma^{*+}_{c}
\Sigma^{+}_{c})$&...&$-$0.11&...&$-$1.2$\pm$0.3&$-$0.101&0.110&$-$0.012&$-$0.003\\
&$\mu(\Sigma^{*0}_{c}
\Sigma^{0}_{c})$&...&1.30&...&0.5$\pm$0.2&1.25&$-$0.04&0.27&1.48\\
&$\mu(\Xi^{*+}_{c}
\Xi^{'+}_{c})$&...&$-$0.26&...&...&$-$0.27&0.18&$-$0.14&$-$0.23\\
&$\mu(\Xi^{*0}_{c}
\Xi^{'0}_{c})$&...&1.11&...&...&1.14&$-$0.04&0.14&1.24\\
&$\mu(\Omega^{*0}_{c} \Omega^{0}_{c})$
&...&0.97&...&...&0.98&$-$0.03&0.01&0.96 \\\hline $6 \to \bar3$ &$
\mu(\Sigma^{*+}_{c}
\Lambda^{+}_{c})$&...&2.33&...&3.8$\pm$1.4&2.30&$-$0.37&0.47&2.40\\
&$\mu(\Xi^{*0}_{c} \Xi^{0}_{c})$
&...&$-$0.29&...&$-$0.45$\pm$0.18&$-$0.29&0.02&$-$0.23& $-$0.50\\
&$\mu(\Xi^{*+}_{c}
\Xi^{+}_{c})$&...&2.14&...&4.0$\pm$1.8&2.20&$-$0.38&0.26&
2.08\\\hline 3 $\to$ 3&$\mu(\Xi^{*++}_{cc} \Xi^{++}_{cc})
$&...&1.42&...&...&1.42&$-$0.37&0.28&1.33\\ &$\mu(\Xi^{*+}_{cc}
\Xi^{+}_{cc}) $&...&$-$1.22&...&...&$-$1.22&0.07&$-$0.26&$-$1.41
\\&$\mu(\Omega^{*+}_{cc} \Omega^{+}_{cc})
$&...&$-$0.91&...&...&$-$0.91&0.03&$-$0.01 &$-$0.89\\\hline
\multicolumn{9}{|c|}{Spin $\frac{1}{2}{^+} \to \frac{1}{2}{^+}$
transitions} \\\hline 8 $\to$ 8 &$ \mu(\Sigma^0
\Lambda)$&1.61$\pm$0.08 \cite{pdg}&1.52&$-1.16 \pm 0.15$& 1.6 $\pm$ 0.3&1.59&$-$0.30&0.31&1.60\\
$\bar3 \to$ 6&$\mu(\Sigma^{+}_{c}
\Lambda^{+}_{c})$&...&1.46&...&1.5$\pm$0.4&1.51&$-$0.24&0.29&
1.56\\
&$\mu(\Xi^{'0}_{c} \Xi^{0}_{c})$&...&$-$0.18&...&...&$-$0.18&0.01&$-$0.14&$-$0.31\\
&$\mu(\Xi^{'+}_{c} \Xi^{+}_{c})$&...&
1.33&...&...&1.37&$-$0.23&0.16& 1.30\\\hline
\end{tabular}
\caption{The baryon magnetic moments for the low lying and charmed
spin $\frac{3}{2}{^+} \to \frac{1}{2}{^+}$ and $\frac{1}{2}{^+} \to
\frac{1}{2}{^+}$ transitions (in units of
$\mu_{N}$).}\label{transnum}
\end{center}
\end{table}


\begin{thebibliography}{99}


\bibitem {ic} S.J. Brodsky, P. Hoyer, C. Peterson, and N. Sakai,
Phys. Lett. {\bf B 93}, 451 (1980); S.J. Brodsky, C. Peterson, and
N. Sakai, Phys. Rev. {\bf D 23}, 2745 (1981).

\bibitem{garcilazo} H. Garcilazo, J. Vijande, and A. Valcarce, J.
Phys. {\bf G 34}, 961 (2007).

\bibitem{pdg} C. Amsler {\it et al.}, Phys. Lett. {\bf B 667}, 1
(2008).

\bibitem{choudhury} A.L. Choudhury and V. Joshi, Phys. Rev. {\bf D
13}, 3115 (1976); {\it ibid.} 3120 (1976); D.B. Lichtenberg, Phys.
Rev. {\bf D 15}, 345 (1977); R.J. Johnson and M. Shah-Jahan, {\it
ibid.} 1400 (1977).

\bibitem{rlpm} S.N. Jena and D.P. Rath, Phys. Rev. {\bf D 34}, 196
(1986).

\bibitem{ccm} L.Ya. Glozman and D.O. Riska, Nucl. Phys. {\bf A
603}, 326 (1996); {\it Erratum-ibid.} {\bf 620}, 510 (1997).

\bibitem{rqm} B. Julia Diaz and D.O. Riska, Nucl. Phys. {\bf A
739}, 69 (2004).

\bibitem{skyrme} Yongseok Oh and Byung-Yoon Park, Mod. Phys.
Lett. {\bf A 11}, 653 (1996).

\bibitem{bsa} S. Scholl and H. Weigel, Nucl. Phys. {\bf A 735}, 163
(2004).

\bibitem{rtqm} A. Faessler {\it et al.}, Phys. Rev. {\bf D 73},
094013 (2006).

\bibitem{patel} S. Kumar, R. Dhir, and R.C. Verma, Jol. Phys. {\bf G
31}, 141 (2005); A. Majethiya, B. Patel, and P.C. Vinodkumar, Eur.
Phys. J.{\bf A 38}, 307 (2008); B. Patel, A.K. Rai, and P.C.
Vinodkumar, Jol. of Phys. {\bf G 35}, 065001 (2008); R. Dhir and
R.C. Verma, Eur. Phys. J. {\bf A 42}, 243 (2009).

\bibitem{wanglee}  F.X. Lee, Phys. Rev. {\bf D 57}, 1801 (1998);
Lai Wang and F.X. Lee, Phys. Rev. {\bf D 78}, 013003 (2008).

\bibitem{qssr} Shi-lin Zhu, W-Y.P. Hwang, and Ze-sen Yang, Phys.
Rev. {\bf D 56}, 7273 (1997).

\bibitem{lcqsr} T.M. Aliev, A. Ozpineci, and M. Savci, Phys. Rev.
{\bf D 66}, 016002 (2002); {\it ibid.} {\bf D 67}, 039901 (2003);
T.M. Aliev, K. Azizi and A. Ozpineci, Phys. Rev. {\bf D 77}, 114006
(2008).

\bibitem{tam3/2} T.M. Aliev, A. Ozpineci, and M. Savci, Phys. Rev.
{\bf D 62}, 053012 (2000).

\bibitem{tam} T.M. Aliev, A. Ozpineci, and M. Savci, Phys. Lett.
{\bf B 516}, 299 (2001); Phys. Rev. {\bf D 65}, 096004 (2002); T.M.
Aliev, K. Azizi, and A. Ozpineci, {\it ibid.} {\bf 79}, 056005
(2009).

\bibitem{charm} T. Hatsuda, and T. Kunihiro, Phys. Rep. {\bf 247},
221  (1994); F.S. Navarra {\it et al.}, Phys. Rev. {\bf D 54}, 842
(1996); A. Blotz and E. Shuryak, Phys. Lett. {\bf B 439}, 415
(1998); M.V. Polyakov, A. Schafer, and O.V. Teryaev, Phys. Rev.
{\bf D 60}, 051502(R) (1999);  M. Franz, M.V. Polyakov, and K.
Goeke, Phys. Rev. {\bf D 62}, 074024 (2000).

\bibitem{wein} S. Weinberg, Physica {\bf A 96}, 327(1979);
A. Manohar and H. Georgi, Nucl. Phys. {\bf B 234}, 189 (1984).

\bibitem{eichten}  E.J. Eichten, I. Hinchliffe, and C. Quigg,
 Phys. Rev. {\bf D 45}, 2269 (1992).

\bibitem{cheng} T.P. Cheng and Ling Fong Li, Phys. Rev. Lett.
{\bf 74}, 2872 (1995);  Phys. Rev. {\bf D 57}, 344 (1998);
hep-ph/9709293.

\bibitem{dgg} A. De Rujula, H. Georgi, and S.L. Glashow, Phys. Rev.
{\bf D 12}, 147 (1975).

\bibitem{isgur} N. Isgur, G. Karl, and R. Koniuk, Phys. Rev. Lett.
{\bf 41}, 1269 (1978); R. Koniuk and N. Isgur, Phys. Rev. {\bf D
21}, 1868 (1980); N. Isgur and G. Karl, Phys. Rev. {\bf D 21},
3175 (1980); N. Isgur {\it et al.}, Phys. Rev. {\bf D 35}, 1665
(1987).

\bibitem{hd} H. Dahiya and M. Gupta, Phys. Rev. {\bf D 64}, 014013
(2001); H. Dahiya and M. Gupta, Int. Jol. of Mod. Phys. A, Vol. 19,
No. 29, 5027 (2004); H. Dahiya, M. Gupta, and J.M.S. Rana, Int. Jol.
of Mod. Phys. A, Vol. 21, No. 21, 4255 (2006); H. Dahiya and M.
Gupta, Eur. Phys. J. {\bf C 52}, 571 (2007).

\bibitem{song} X. Song, J.S. McCarthy, and H.J. Weber, Phys. Rev.
{\bf D 55}, 2624 (1997); X. Song, Phys. Rev. {\bf D 57}, 4114
(1998).

\bibitem{hds} H. Dahiya and M. Gupta, Phys. Rev. {\bf D 78}, 014001
(2008).

\bibitem{ns} N. Sharma, H. Dahiya, P.K. Chatley, and M. Gupta,
 Phys. Rev. {\bf D 79}, 077503 (2009).

\bibitem{cgsr} S. Coleman and S.L. Glashow, Phys. Rev. Lett. {\bf
6}, 423(1961).


\bibitem{cheng1} T.P. Cheng and Ling Fong Li, Phys. Rev. Lett. {\bf
80}, 2789 (1998).

\bibitem{johan} J. Linde, T. Ohlsson, and H. Snellman, Phys. Rev.
{\bf D 57}, 452 (1998); {\it ibid.} {\bf 57}, 5916 (1998).

\bibitem{hdorbit} H. Dahiya and M. Gupta, Phys. Rev. {\bf D 66},
051501(R) (2002);  H. Dahiya and M. Gupta, Phys. Rev. {\bf D 67},
114015 (2003).

\bibitem{chengspin} T.P. Cheng and Ling Fong Li, hep-ph/9811279.

\bibitem{songcharm} X. Song, Phys. Rev. {\bf D 65}, 114022 (2002);
X. Song, Int. Jol. of Mod. Phys. {\bf A 18}, 11501 (2003).

\bibitem{hdcharm} H. Dahiya and M. Gupta, Phys. Rev. {\bf D
67}, 074001 (2003).

\bibitem{emc} J. Ashman {\it et al.} (European Muon Collaboration),
Phys. Lett. {\bf B 206}, 364 (1988); Nucl. Phys. {\bf  328}, 1
(1989).

\bibitem{adams} B. Adeva {\it et al.} (SMC Collaboration), Phys.
Lett. {\bf B 302}, 533 (1993); D. Adams {\it et al.}, Phys. Rev.
{\bf D 56}, 5330 (1997).

\bibitem{ellis} J. Ellis and M. Karliner, Phys. Lett. {\bf B 313},
131 (1993); {\it ibid.} {\bf 41}, 397 (1995).

\bibitem{nmc} P. Amaudruz {\it et al.} (New Muon Collaboration),
Phys. Rev. Lett. {\bf 66}, 2712 (1991); M. Arneodo {\it et al.},
Phys. Rev. {\bf D 50}, R1 (1994).

\bibitem{e866} E.A. Hawker {\it et al.} (E866/NuSea Collaboration),
Phys. Rev. Lett. {\bf 80}, 3715 (1998); J.C. Peng {\it et al.},
Phys. Rev. {\bf D 58}, 092004 (1998); R.S. Towell {\it et al.},
Phys. Rev. {\bf D 64}, 052002 (2001).

\bibitem{yaouanc} A. Le Yaouanc, L. Oliver, O. Pene, and J.C. Raynal,
Phys. Rev. {\bf D 12}, 2137 (1975); {\it ibid.} {\bf 15}, 844
(1977); M. Gupta and A.N. Mitra, Phys. Rev. {\bf D 18}, 1585 (1978).

\bibitem{mpi1} V. Elias, Mong Tong, and M.D. Scadron, Phys. Rev. {\bf
D 40}, 3670 (1989); Duane A. Dicus, Djordge Minic, Ubirajara van
Klock, and Roberto Vega, Phys. Lett. {\bf 284}, 384 (1992); Y.B.
Dong{\it et al.}, J. Phys. {\bf G 25}, 1115 (1999).

\bibitem{effm} Ikuo S. Sogami and Noboru Oh\'yamaguchi, Phys.
Rev. Lett. {\bf 54}, 2295 (1985); Kuang-Ta Chao, Phys. Rev. {\bf D
41}, 920 (1990); M. Gupta, J. Phys. {\bf G 16}, L213 (1990).

\bibitem{conf} B.O. Kerbikov and Yu. A. Simonov, Phys. Rev. {\bf D
62}, 093016 (2000); B.O. Kerbikov, Phys. Atom. Nucl. {\bf 64}, 1856
(2001).

\bibitem{lattice} D.B. Leinweber, R.M. Woloshyn, and T. Draper,
Phys. Rev. {\bf D 43}, 1659 (1991); D.B. Leinweber, Phys. Rev. {\bf
D 45}, 252 (1992); I.C. Cloet, D.B. Leinweber, and A.W. Thomas,
Phys. Lett. {\bf B 563} 157 (2003); D.B. Leinweber, T. Draper, and
R.M. Woloshyn, Phys. Rev. {\bf D 46}, 3067 (1992).

\bibitem{kotulla} M. Kotulla {\it et al.}, Phys. Rev. Lett. {\bf
89}, 272001 (2002); M. Kotulla, Prog. Part. Nucl. Phys. {\bf 61},
147 (2008).

\bibitem{diehl} H.T. Diehl {\it et al.}, Phys. Rev. Lett. {\bf 67},
804 (1991).

\bibitem{tiator} L. Tiator {\it et al.}, Nucl. Phys. {\bf A
689}, 205 (2001).

\bibitem{yaoubook}  A. Le Yaouanc {\it et al.}, {\it Hadron
Transitions in the Quark Model}, Gordon and Breach (1988).



\end{thebibliography}
\end{document}